\documentclass[twocolumn]{aastex62}
\usepackage{amsmath}
\graphicspath{}

\newcommand{\acronym}[1]{{\small{#1}}}

\newcommand{\package}[1]{\textsl{#1}}
\newcommand{\gaia}{\textsl{Gaia}}

\newcommand{\decam}{DECam}
\newcommand{\DR}[1]{\acronym{DR#1}}

\newcommand{\bs}[1]{\boldsymbol{#1}}

\newcommand{\articlename}{\textsl{Article}}
\newcommand{\sectionname}{Section}
\newcommand{\equationname}{Equation}

\newcommand{\given}{\,|\,}
\newcommand{\norm}{\mathcal{N}}

\newcommand{\dd}{\mathrm{d}}

\newcommand{\mean}[1]{\left< #1 \right>}
\newcommand{\mat}[1]{\mathbf{#1}}

\newcommand{\msun}{\textrm{M}_\odot}
\newcommand{\kpc}{\textrm{kpc}}
\newcommand{\kms}{\ensuremath{\textrm{km}~\textrm{s}^{-1}}}
\newcommand{\masyr}{\ensuremath{\textrm{mas}~\textrm{yr}^{-1}}}

\newcommand{\feh}{\ensuremath{[\textrm{Fe} / \textrm{H}]}}

\newcommand{\hi}{H{\footnotesize I} }

\newcommand{\clustername}{\textsl{Price-Whelan 1}}
\newcommand{\lmcsmc}{LMC--SMC}
\newcommand{\bprp}{\ensuremath{\textrm{BP} - \textrm{RP}}}
\newcommand{\truepm}{\ensuremath{\tilde{\bs{\mu}}}}

\newcommand{\eep}{\ensuremath{\acronym{\textrm{EEP}}}}
\newcommand{\Nisofit}{417}

\newcommand{\clage}{\ensuremath{117~\textrm{Myr}}}
\newcommand{\clfeh}{\ensuremath{-1.14}}
\newcommand{\cldist}{\ensuremath{28.7~\textrm{kpc}}}

\newcommand{\clageerr}{\ensuremath{117 \pm 23~\textrm{Myr}}}
\newcommand{\clfeherr}{\ensuremath{-1.14 \pm 0.05}}
\newcommand{\cldisterr}{\ensuremath{28.7 \pm 0.5~\textrm{kpc}}}
\newcommand{\claverr}{\ensuremath{0.205 \pm 0.011}}

\newcommand{\changes}[1]{#1}
\newcommand{\changestwo}[1]{#1}
\newcommand{\changesthr}[1]{#1}

\shorttitle{A recent star formation event in the Magellanic stream}
\shortauthors{Price-Whelan et al.}

\begin{document}

\title{\textbf{
Discovery of a disrupting open cluster far into the Milky Way halo:\\
a recent star formation event in the leading arm of the Magellanic stream?}}

\author[0000-0003-0872-7098]{Adrian~M.~Price-Whelan}
\affiliation{Department of Astrophysical Sciences,
             Princeton University,
             4 Ivy Lane, Princeton, NJ 08544, USA}
\email{adrn@astro.princeton.edu}
\correspondingauthor{Adrian M. Price-Whelan}

\author[0000-0002-1793-3689]{David~L.~Nidever}
\affiliation{Department of Physics, Montana State University, P.O. Box 173840, Bozeman, MT 59717, USA}
\affiliation{National Optical Astronomy Observatory, 950 North Cherry Ave, Tucson, AZ 85719, USA}

\author[0000-0003-1680-1884]{Yumi~Choi}
\affiliation{Department of Physics, Montana State University, P.O. Box 173840, Bozeman, MT 59717, USA}
\affiliation{Steward Observatory, University of Arizona, 933 North Cherry Avenue, Tucson AZ, 85721, USA}

\author[0000-0002-3569-7421]{Edward~F.~Schlafly}
\affiliation{Lawrence Berkeley National Laboratory, One Cyclotron Road, Berkeley, CA 94720, USA}


\author[0000-0002-8537-5711]{Timothy~Morton}
\affiliation{Department of Astrophysical Sciences,
             Princeton University,
             4 Ivy Lane, Princeton, NJ 08544, USA}
\affiliation{Department of Astronomy, University of Florida, 211 Bryant Space Science Center, P.O Box 112055, Gainesville, FL 32611-2055 USA}
\affiliation{Center for Computational Astrophysics, Flatiron Institute, 162 5th Avenue, New York, NY 10010, USA}

\author[0000-0003-2644-135X]{Sergey E. Koposov}
\affiliation{McWilliams Center for Cosmology, Carnegie Mellon University, 5000 Forbes Avenue, Pittsburgh, PA 15213, USA}
\affiliation{Institute of Astronomy, University of Cambridge, Madingley Road, Cambridge CB3 0HA, UK}

\author[0000-0002-0038-9584]{Vasily Belokurov}
\affiliation{Institute of Astronomy, University of Cambridge, Madingley Road, Cambridge CB3 0HA, UK}
\affiliation{Center for Computational Astrophysics, Flatiron Institute, 162 5th Avenue, New York, NY 10010, USA}

\begin{abstract}

\changes{We report the discovery of a young ($\tau \sim \clage$), low-mass ($M \sim 1200~\msun$), metal-poor ($\feh \sim \clfeh$) stellar association at a heliocentric distance $D \approx \cldist$, placing it far into the Milky Way halo.}
At its present Galactocentric position $(R, z) \sim (23, 15)~\textrm{kpc}$, the association is (on the sky) near the leading arm of the gas stream emanating from the Magellanic cloud system, but is located $\approx 60^\circ$ from the Large Magellanic Cloud (LMC) center on the other side of the Milky Way (MW) disk.
If the cluster is co-located with \hi gas in the stream, we directly measure the distance to the leading arm of the Magellanic stream.
The measured distance is inconsistent with Magellanic stream model predictions that do not account for ram pressure and gas interaction with the MW disk.
The estimated age of the cluster is consistent with the time of last passage of the leading arm gas through the Galactic midplane; We therefore speculate that this star-formation event was triggered by its last disk midplane passage. 
Most details of this idea remain a puzzle: the Magellanic stream has low column density, the MW disk at large radii has low gas density, and the relative velocity of the leading arm and MW gas is large.
However it formed, the discovery of a young stellar cluster in the MW halo presents an interesting opportunity for study.
This cluster was discovered with \gaia\ astrometry and photometry alone, but follow-up \decam\ photometry was crucial for measuring its properties.
\end{abstract}


\section{Introduction} \label{sec:intro}

The \changes{distant stellar} halo of the Milky Way (i.e. far from the disk midplane, $|z| \gtrsim 5~\kpc$) is typically characterized by its old ($\gtrsim 10~\textrm{Gyr}$), metal-poor ($\feh \approx -1.5$) stellar population.
The typical old age of the stellar halo is understood as a signature of the dominant (in stellar mass) progenitor systems that were accreted early on in the formation of the Galaxy \citep[massive dwarf galaxies][]{Deason:2015, Fiorentino:2015}.
It is thought that these systems came in with significant gas reservoirs, but were quenched and stripped through collisional processes that heated and dispersed the gas \citep[e.g.,][]{Mayer:2006}, thus preventing immediate star formation in the deposited gas.
The Milky Way, however, continues to accrete satellite galaxies, as is evidenced by the prominent stellar stream from the Sagittarius dwarf galaxy \citep{Ibata:1994, Majewski:2003}, the presence of the Large and Small Magellanic Clouds (LMC and SMC), and about 50 dwarf satellites within the Galactic halo.
While Sagittarius was likely stripped of its neutral gas long ago \citep{Burton:1999, Tepper-Garcia:2018}, the \lmcsmc\ system is associated with $\approx 8\times 10^8~\msun$ of \hi gas \citep{Bruns:2005}, which extends into leading and trailing gas streams \citep{Mathewson:1974, Putman:1998, Bruns:2005, Nidever:2010}.
Here we report the first discovery of a young star cluster in the Milky Way halo that appears to be associated with this \lmcsmc\ gas stream, which suggests that young stars can form from tidally stripped gas during low-mass mergers and may therefore exist throughout the otherwise aging stellar halo.

The Magellanic stream (MS), including gas in the leading arm (LA), is a large stream of predominantly hydrogen gas emanating from the \lmcsmc\ system that wraps nearly $\approx 200^\circ$ around the sky \citep{Mathewson:1974, Putman:1998, Bruns:2005, Nidever:2010} and contains a significant fraction of the total gas mass associated with the \lmcsmc\ \citep{Bruns:2005}.
The trailing MS has been studied in great detail by large-area and high-resolution radio sky surveys: Recent surveys have found small-scale structure and gas fragmentation \citep[e.g.,][]{Nidever:2008, For:2014} and a large-scale bifurcation, with kinematically \citep{Nidever:2008} and chemically \citep{Fox:2013} distinct ``strands'' that lead back to the LMC and the SMC.
The LA gas has been found to connect to regions of low-column-density gas ($N\sim 10^{18}$--$10^{19}~\textrm{cm}^{-2}$) on the other side of the Galactic disk \citep{Putman:1998, Nidever:2008}, and has been decomposed into distinct gas features named LA I--IV \citep{Bruns:2005, Nidever:2008, Venzmer:2012}.

The origin and formation of these LA features is still uncertain.
Initial studies of the LA argued that the features closest to the \lmcsmc\ can be traced back to the SMC \citep{Putman:1998}, but outer features of the LMC appear to lead directly into the LA I feature \citep{Nidever:2008}.
Recent chemical abundance measurements along several sight-lines passing through the LA again support an SMC origin for the gas \citep{Fox:2013, Fox:2018, Richter:2018}.
Whatever the origin of the LA gas, it is clear that tidal stripping by the Milky Way is required to form the LA \citep{Nidever:2008, Besla:2012}.
However, the LA features deviate from the predicted orbit of the \lmcsmc, implying that ram pressure or interactions the outer Milky Way disk (from the recent disk plane passage) may have removed orbital energy from the LA gas \citep[e.g.,][]{Bekki:2008}.

The gas in the MS encodes information about the past and future trajectory of the \lmcsmc, and about interactions between this gas and the Milky Way.
Combined with recent proper motion measurements of the LMC \citep{Kallivayalil:2006, Kallivayalil:2013} and improved models for the \lmcsmc\ that suggest they are on their first passage through the Galaxy \citep{Besla:2007, Besla:2010, Besla:2012}, several groups have used the MS to constrain properties of both the interaction history of the \lmcsmc\ and of the dark matter halo of the Milky Way (see recent review by \citealt{DOnghia:2016}).
In general, these models for the formation of the MS rely on past interactions between the LMC and SMC to preprocess the Magellanic gas distribution before infall and eventual stripping by the tidal field of the Milky Way \citep{Besla:2012, Diaz:2012};
More recent hydrodynamical simulations of the MCs showed that the repeated encounters between the \lmcsmc\ strip gas both from the SMC and the LMC, and this tidally stripped gas can then create filamentary structures both in the leading and trailing MS \citep{Pardy:2018}.
Many observational studies \citep[e.g.,][]{Olsen:2011, Noel:2013, Mackey:2016, Carrera:2017, Choi:2018a, Choi:2018b, Zivick:2018, Belokurov:2019} of the \lmcsmc\ themselves will provide strong constraints the interaction history of the MCs.

One critical difficulty in using the MS to further improve models of the \lmcsmc\ interaction and infall is the lack of distance information along the MS.
No significant over-density of stars have been found associated with the trailing MS \citep{Guhathakurta:1998}, thus leaving distance and tangential velocity information unknown.
While a small number of OB stars have been found in the vicinity of the LA gas \citep{Casetti-Dinescu:2014, Zhang:2017}, their sparsity and concentration near the Galactic plane make it difficult to unambiguously associate them with the MS and not runaway OB stars from the Milky Way disk.

In this \articlename, we report the discovery of a young stellar cluster at the far edge of the LA II feature ($L_{\textrm{MS}} \sim 65^\circ$) that is located far into the Galactic halo ($D \sim \cldist$) and therefore plausibly formed from gas in the leading arm of the MS as it crossed the Galactic disk.
This provides the first precise distance measurement to the MS leading arm to be compared with simulations of the \lmcsmc\ in the Milky Way, and provides an opportunity to study recent star formation in a unique environment.
The discovery of \clustername\ will enable new modeling efforts that track the infall of the \lmcsmc, the tidal stripping of Magellanic gas, and the interaction of this gas with the Milky Way.

In \sectionname~\ref{sec:data}, we present the initial discovery with \gaia\ \DR{2} and follow-up observations with \decam\ to obtain deeper photometry of the region around the association.
In \sectionname~\ref{sec:methods}, we use the \gaia\ data to measure the kinematics, and \decam\ photometry to infer the age, metallicity, and distance to \clustername
In \sectionname~\ref{sec:results}, we interpret the inferred stellar population parameters of \clustername\ and discuss plausible formation scenarios.
We conclude in \sectionname~\ref{sec:conclusion}.

\begin{figure*}[t!]
\centering
\includegraphics[width=\textwidth]{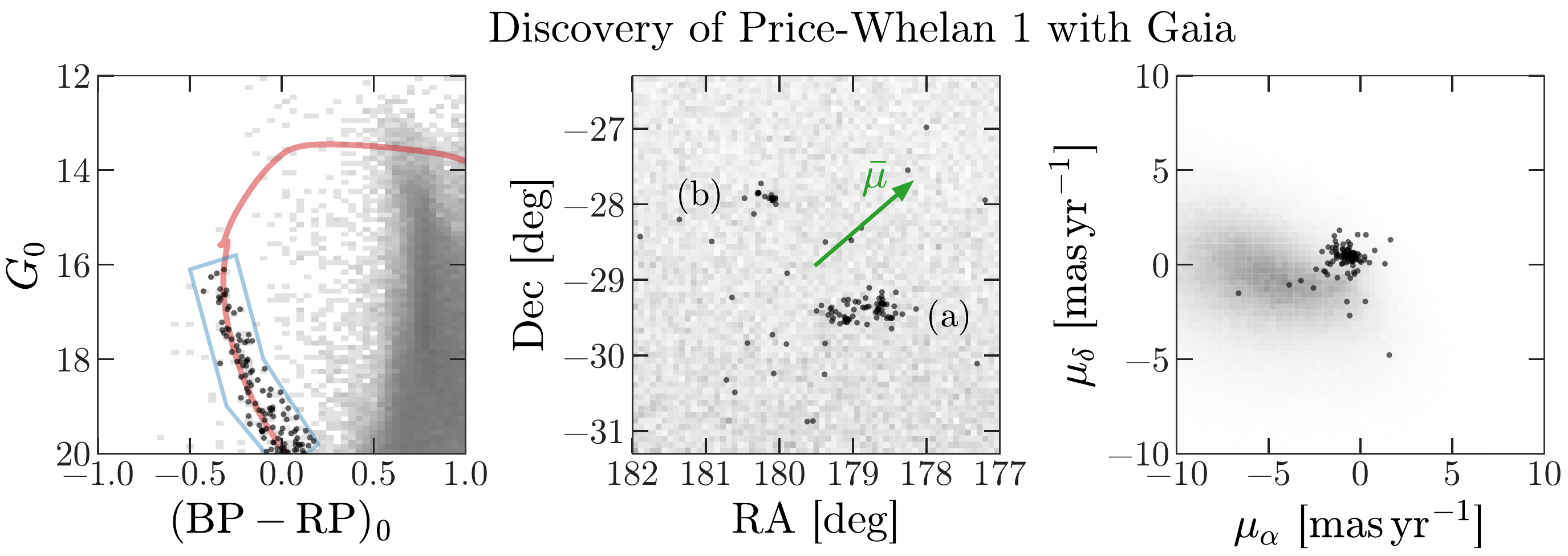}
\caption{Color-magnitude diagram, sky positions, and proper motions from \gaia\ \DR{2} for the region around $(\alpha, \delta) \sim (179, -29)^\circ$.
\textbf{Left panel}: \gaia\ color-magnitude diagram (CMD) for all sources with $G < 20$ and sky position within 3 degrees of $(\alpha, \delta) = (179.5, -28.8)^\circ$, extinction corrected following \citet{Danielski:2018}.
Red line shows a $100~\textrm{Myr}$, $\feh = -1.1$ \acronym{MIST} isochrone shifted to a distance of $30~\kpc$.
Shaded pixels (grey 2D histogram) show the density of all sources in the CMD for this sky region, and points (black markers) show only sources in the blue selection polygon shown.
\textbf{Middle panel}: Sky positions for all sources in the shown sky region (grey 2D histogram).
Black markers show only sources in the blue selection polygon shown in the CMD (left panel).
Arrow (green) indicates the inferred proper motion direction (\sectionname~\ref{sec:pmclean}).
Subcomponents of the cluster (a and b) are indicated.
\textbf{Right panel}: Proper motions for all sources in the shown sky region (grey 2D histogram).
Black markers again show only sources in the blue CMD selection region (left panel).
}
\label{fig:cmds}
\end{figure*}

\section{Data} \label{sec:data}

\subsection{Cluster discovery with \gaia}
\label{sec:discovery}

We use astrometric data from the \gaia\ mission (\citealt{Prusti:2016}), data release 2 (\DR{2}; \citealt{Gaia-Collaboration:2018, Lindegren:2018}) to search for distant, comoving multiplets of blue stars.
Our original intent was to search for small, distant, comoving groups of blue horizontal branch stars to identify new candidate satellites of the Milky Way.
We therefore initially select all stars from \gaia\ with parallax $\varpi < 1$, color $-0.5 < (\bprp) < 0$, $G$-band magnitude $G < 20$, and Galactic latitude $|b| > 20^\circ$ (see Appendix~\ref{sec:queries} for the database query).
We further exclude stars within a $15^\circ$ radius from the LMC, and a $8^\circ$ radius from the SMC --- 27,895 stars remain after these cuts.
We then cross-match this catalog to itself with both sky positions and proper motions: we search for pairs of stars that have separations $s < 0.5^\circ$ and proper motion differences $|\Delta \mu| < 0.5~\masyr$.
We then combine mutually-connected comoving pairs into small groups of stars that are colocated on the sky and comoving in proper motions, and remove groups that have $<4$ members.
We cross-match the mean sky positions of the groups to locations of local group galaxies \citep{McConnachie:2012} and Milky Way globular clusters \citep[2010 edition;][]{Harris:1996} and filter out all groups that lie within 1 degree of these known objects.
After these filters, one group of comoving stars remains at $(\textrm{RA}, \textrm{Dec}) \sim (179, -29)^\circ$.

We then query all objects from the \gaia\ \DR{2} catalog within a rectangle centered on the nominal position of this group, with a width of $5^\circ$ and a height of $5^\circ$ in the equatorial (ICRS) coordinate system (see Appendix~\ref{sec:queries} for the database query).
\figurename~\ref{fig:cmds} shows the \gaia\ data for this region:
The left panel shows the \gaia\ color-magnitude diagram, extinction corrected following the procedure used in \citet{Danielski:2018} \changes{and the coefficients from \citet{Babusiaux:2018}}, with the blue over-density highlighted by the polygon (blue) and under-plotted with a $100~\textrm{Myr}$, $\feh = -1.1$ \acronym{MIST} isochrone \citep[red line;][]{Dotter:2016, Choi:2016, Paxton:2011, Paxton:2013, Paxton:2015}.
Surprisingly, this group appears to be a young, distant main sequence, rather than an old population of horizontal branch stars.
The middle and right panels of \figurename~\ref{fig:cmds} show sky positions and proper motions of all stars in this sky region (grey background density), and only stars in the CMD selection polygon (black markers).

The \gaia\ data reveal the presence of a young, distant, spatially-clustered, and co-moving stellar over-density --- named \clustername\ --- but the \gaia\ photometry is too shallow to resolve anything but the brightest main sequence stars.
The spatial morphology of the cluster is large on the sky, and interestingly substructured with at least two subcomponents (labeled a and b in \figurename~\ref{fig:cmds}) that are indistinguishable in terms of their proper motions and uncertainties.
With the \gaia\ data alone, the distance, age, and metallicity of the cluster  cannot be determined, as these quantities are degenerate where the main sequence is nearly vertical.
In the next section, we describe deeper \decam\ imaging obtained over a portion of the cluster.

\subsection{Follow-up with \decam}
\label{sec:decam}

\begin{figure}[t!]
\centering
\includegraphics[width=0.45\textwidth]{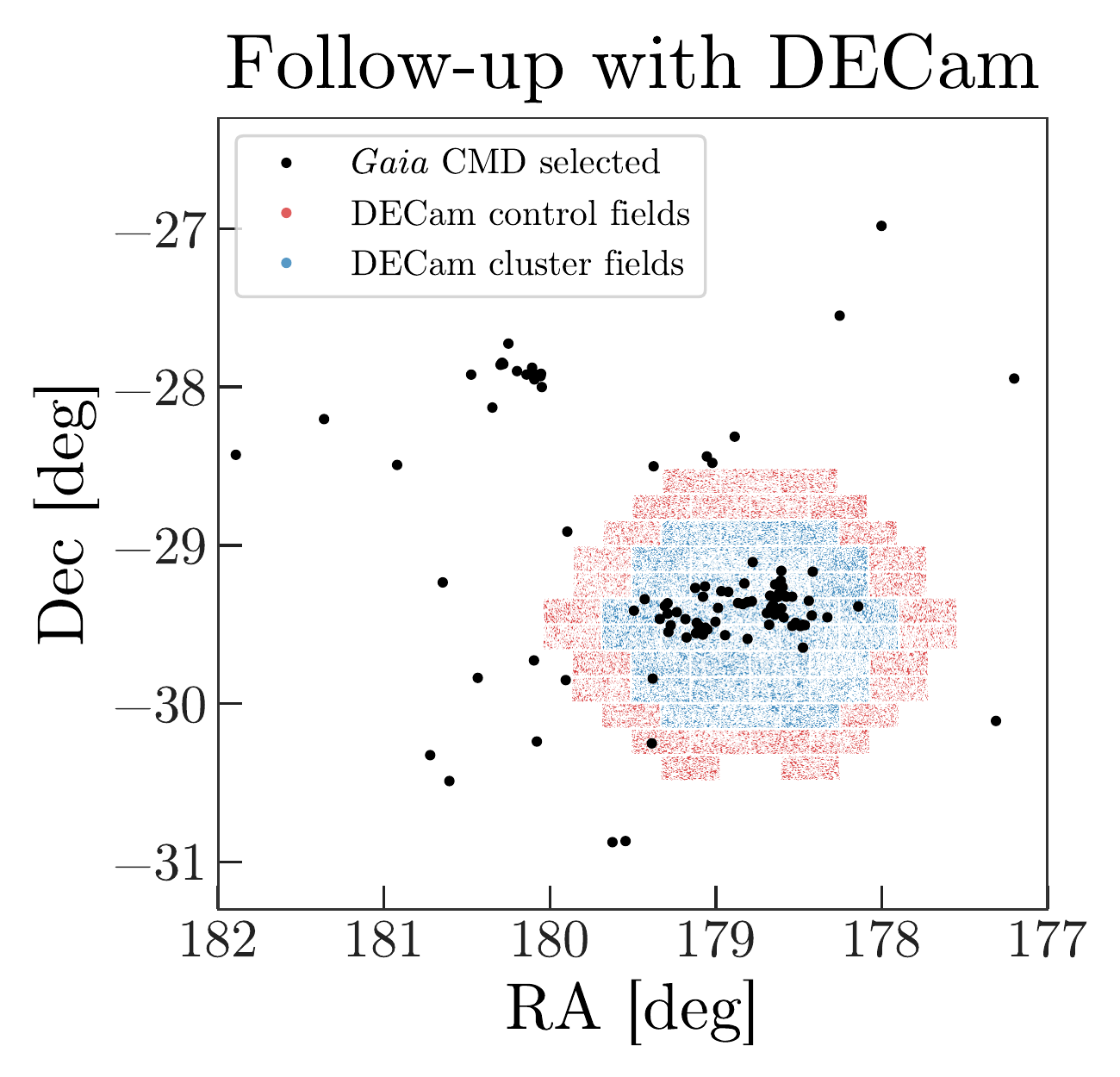}
\caption{The sky region around \clustername, showing the same \gaia\ CMD-selected sources as in \figurename~\ref{fig:cmds} (black markers), and point sources identified from \decam\ g-band follow up (blue and red markers).
Red markers show sources in the \decam\ field used as control sources, and blue markers show sources used as cluster member candidates.}
\label{fig:decam-field}
\end{figure}

We obtained \decam\ u-, g-, and i-band imaging of a single field centered on the ``a'' spatial component of the cluster (see \figurename~\ref{fig:cmds}) discovered using the \gaia\ data (see previous section).  Observations were obtained with the Dark Energy Camera (DECam) on the CTIO 4m Blanco telescope on UT 2018 May 20 with 3$\times$300s
exposures in $u-$, $g-$, and $i$-band.  The NOAO Community Pipeline (CP; \citealt{Valdes:2014}; Valdes et al., in preparation) InstCal images, which have the instrumental signature removed, were downloaded from the NOAO Archive\footnote{\url{http://archive.noao.edu}} for further processing.  Forced PSF photometry was performed with the PHOTRED pipeline \citep{Nidever:2017} which uses the DAOPHOT ALLSTAR \citep{Stetson:1987} and ALLFRAME \citep{Stetson:1994} suite of programs.

\changes{The instrumental photometry (i.e., $-2.5\log(\textrm{ADU}~\textrm{sec}^{-1}$) was calibrated using Pan-STARRS1 \citep[PS1;][]{Chambers:2016} photometry for g- and i-band and with SkyMapper \citep{Wolf:2018} for the u-band observations.
We use a robust linear fit to cross-matched sources (that appear in both PS1/SkyMapper and our observations) to obtain a zero-point and (for g- and i-band) color-term to derive color transformations relating our instrumental DECam photometry to calibrated PS1 grizy and SkyMapper u-band photometry (see \tablename~\ref{tbl:transphot} for the derived transformation coefficients).
We apply this color transformation to the instrumental photometry to put our photometry on the PS1/DECam system. 
This calibration enables us to treat the photometry as if it were PS1 (g- and i-band) and DECam (u-band) photometry in the isochrone modeling described below.
The robust RMS around the linear fit is 0.028 mag for g- and i-band and 0.164 mag for the u-band; We adopt these RMS values as systematic errors that are added in quadrature to the photometric uncertainties for individual sources in the modeling described below.
The larger scatter in u-band---larger than the uncertainty in the individual photometric measurements---is due to the significant differences between the DECam and SkyMapper u-band passbands as well as the u-band's sensitivity to both temperature and metallicity.
}

\figurename~\ref{fig:decam-field} shows the sky positions of the \gaia\ CMD-selected sources (black markers; see same in middle panel, \figurename~\ref{fig:cmds}), along with point sources identified in the g-band data obtained with \decam: sources in control fields are shown as red markers, and sources in cluster fields are shown as blue markers.
We note that the smaller over-density located northeast of the \decam\ field, component b (see \figurename~\ref{fig:cmds}), was not followed up in this work.

\begin{figure}[t!]
\centering
\includegraphics[width=0.45\textwidth]{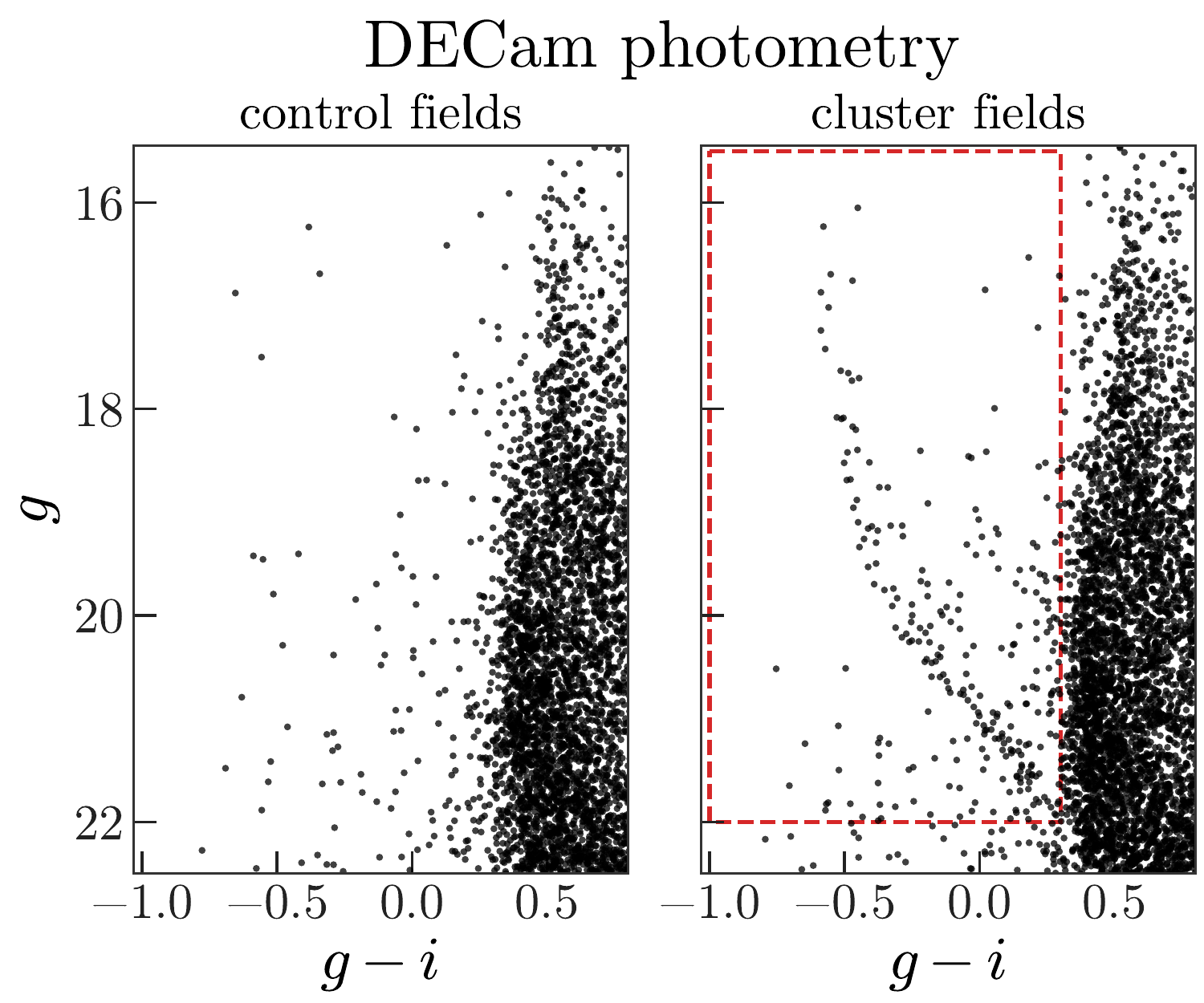}
\caption{\decam\ $g-i$ versus g color-magnitude diagrams for the \decam\ control fields (left panel) and \decam\ cluster fields (right panel).
Note the prominent young, distant main sequence in the cluster fields: This is the main sequence of \clustername.
The sources in the dashed (red) rectangle in the right panel are later used to measure the cluster stellar population parameters.
}
\label{fig:decam-cmd}
\end{figure}

\figurename~\ref{fig:decam-cmd} shows \decam\ color-magnitude diagrams for the control and cluster sub-fields (the magnitudes here are not extinction-corrected).
The blue comoving group, \clustername, identified in \sectionname~\ref{sec:discovery} using \gaia\ data alone shows up as a clear young main sequence in the cluster fields (right panel of \figurename~\ref{fig:decam-cmd}).
Later, we use photometry in the sub-region of the cluster fields identified by the dashed rectangle to infer the cluster parameters.

\begin{center}
\begin{deluxetable}{ccrrr}
\tablecaption{DECam Photometric Transformation\label{tbl:transphot}}
\tablecolumns{5}
\tablewidth{400pt}
\tablehead{
\colhead{Band} & \colhead{Color} & \colhead{Zero-point term} & \colhead{Color term}
}
\startdata
$u_{\rm DECam}$  &  \nodata & $27.178 \pm  0.016$ &     \nodata \\
$g_{\rm PS1}$  & $g_{\rm inst.}-i_{\rm inst.}$ & $24.9471 \pm  0.0068$  &     $-0.03494 \pm 0.0051$ \\
$i_{\rm PS1}$   &  $g_{\rm inst.}-i_{\rm inst.}$ & $24.7534 \pm 0.0011$ &     $-0.07379 \pm 0.0007$
\enddata
\end{deluxetable}
\end{center}

\section{Methods} \label{sec:methods}

In the subsections below, we perform two independent analyses of the data available for \clustername.
First, in \sectionname~\ref{sec:pmclean}, we use astrometric data from \gaia\ \DR{2} to determine the mean proper motion of the cluster by modeling the kinematics of the cluster and background sources.
Then, in \sectionname~\ref{sec:popmodel}, we use photometric data from \decam\ to assign membership probabilities to sources and simultaneously measure the cluster stellar population parameters (age, metallicity, distance, etc.).
We perform these analyses separately because \gaia\ \DR{2} only includes the most massive members of the cluster, but the addition of lower main sequence members apparent in the \decam\ photometry provide a much better constraint on the cluster parameters.
The parameters derived from this analysis are summarized in \tablename~\ref{tbl:clusterparams}.

\begin{figure*}[t!]
\centering
\includegraphics[width=0.9\textwidth]{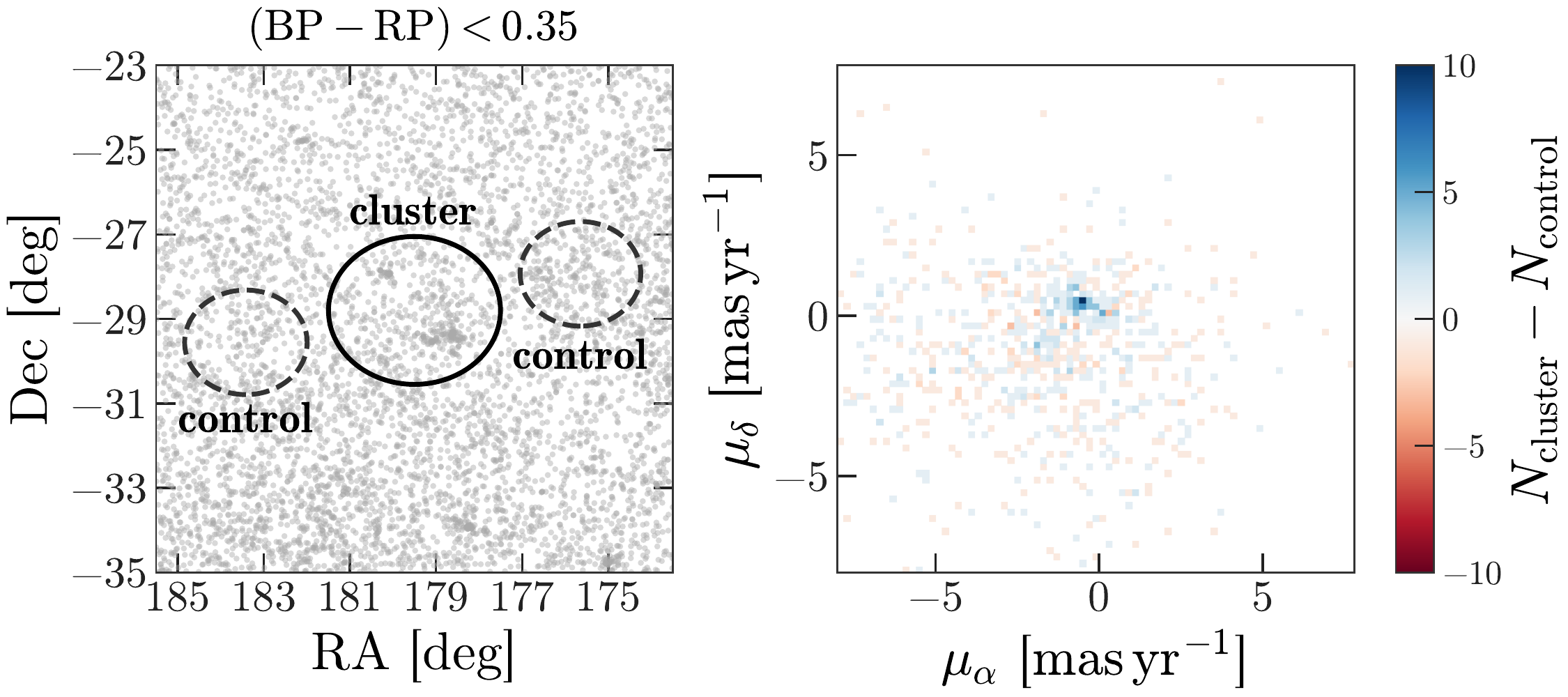}
\caption{\textbf{Left panel}: Sky positions of all \gaia\ sources in this region with $\bprp < 0.35$ (grey markers).
The circles show the regions used as cluster and control fields in the proper motion inference (\sectionname~\ref{sec:pmclean}).
The control fields, together, have the same area as the cluster field.
\textbf{Right panel}: The colored 2D density shows the difference in number of sources per pixel in the cluster field versus the two control fields (see left panel).
The clear over-density in the cluster field (blue) are stars in \clustername.}
\label{fig:pm-members}
\end{figure*}

\subsection{Inferring the mean proper motion with \gaia}
\label{sec:pmclean}

We measure the mean proper motion of \clustername\ by constructing a probabilistic model of the cluster and background populations using astrometric data from \gaia.
We start by selecting all stars with $(\bprp)_0 < 0.35$ to remove low-mass and old main sequence star contamination in the region.
\figurename~\ref{fig:pm-members}, left (grey points), shows the sky positions of stars that pass this blue cut in the region around the young cluster.
The larger solid-line circle indicates the region we define as the cluster area, and the two smaller dashed-line circles indicate control fields that are combined and used for modeling the background distribution of proper motions.
The control fields are designed to, together, have the same total area as the cluster field, and were chosen to have similar latitudes as the cluster field in the Magellanic stream coordinate system \citep{Nidever:2008}.
We assume that the background density in the cluster field is equivalent to the average background density of the joint control fields: \changesthr{We do not see any gradients or significant differences in the proper motion distribution between these two control fields.}
From visual inspection of the CMDs of the control fields, we do not see any significant clustered over-density and therefore assume that these fields are dominated by the background stellar density.
The right panel of \figurename~\ref{fig:pm-members} shows the difference of the 2D proper motion distributions in the cluster and summed control fields.
The distinct over-density of stars near $(\mu_\alpha, \mu_\delta) \approx (-0.5, 0.5)~\masyr$ is the identified comoving association of blue stars.\footnote{Throughout this article, we use $\mu_\alpha$ to refer to the proper motion value provided by \gaia, which includes the $\cos\delta$ term.}
Note that \figurename~\ref{fig:pm-members} is only meant as an illustration: The cluster proper motion is determined probabilistically by taking into account the full \gaia\ covariance matrices for each source, as described below.

To measure the cluster (mean) proper motion, we first construct a model for the error-deconvolved proper motion distribution in the control fields using ``extreme deconvolution'' \citep[XD;][]{Bovy:2011} with two Gaussian components. XD takes into account the full error distributions for each proper motion measurement $\bs{\mu} = (\mu_\alpha, \mu_\delta)$, including covariances $\mat{C}_\mu$, provided by \gaia\ \DR{2}.
After running XD on the proper motion distribution of the control fields, we fix the parameters of the density model and use this as the background model for the cluster field.
We model the proper motion distribution in the cluster region using a two-component mixture model with a single, isotropic Gaussian component for the error-deconvolved cluster distribution with mean $\bs{x}$ and isotropic variance $s^2$, and the XD-inferred background model, $p_{\textrm{XD}}$, for the background component.
In detail, taking $f$ to be the fraction of blue stars in this region belonging to the young cluster, and \truepm\ to be the true proper motion for a single star,
\begin{align}
    p(\bs{\mu} \given \truepm, \mat{C}_\mu) &=
        \norm(\bs{\mu} \given \truepm, \mat{C}_\mu)\\
    p(\truepm \given f, \bs{x}, s) &=
        f \, p_{\textrm{cl}}(\truepm \given \bs{x}, s)
        + (1-f) \, p_{\textrm{XD}}(\truepm)\\
    p_{\textrm{cl}}(\truepm \given \bs{x}, s) &=
        \norm(\truepm \given \bs{x}, s^2 \, \mathbb{I})
\end{align}
where $\norm(\cdot \given \bs{y}, \mat{C})$ represents the multidimensional normal distribution with mean $\bs{y}$ and covariance matrix $\mat{C}$, and $\mathbb{I}$ is the identity matrix.
Because all distributions are Gaussian, the per-star parameters (the true proper motions, \truepm) can be analytically marginalized out so that the per-star likelihood can be expressed as $p(\bs{\mu} \given f, \bs{x}, s, \mat{C}_\mu)$.
We assume that the measurements for each star, $n$, are independent so that the full likelihood of all $N$ stars given the parameters $(f, \bs{x}, s)$ is
\begin{align}
    p(\{\bs{\mu}_n\}_N \given f, \bs{x}, s, \{\mat{C}_{\mu, n}\}_N) &=
        \prod_n^N p(\bs{\mu}_n \given f, \bs{x}, s, \mat{C}_{\mu, n}) \quad .
        \label{eq:likelihood}
\end{align}

We use an ensemble Markov chain Monte Carlo (MCMC) sampler \citep[\texttt{emcee};][]{emcee, Goodman:2010} to generate posterior samples over the parameters $(f, \bs{x}, s)$ using the likelihood defined above (\equationname~\ref{eq:likelihood}), and assuming the following prior probability distributions: uniform over the domain $(-5, 5)~\masyr$ for each component of $\bs{x}$, uniform in $f$, and uniform in log-$s$ over the domain $-6 < \ln s < 4$ (with $s$ in units of \masyr).
We run the sampler with 32 walkers for 256 steps as burn-in, then reset the sampler and run for an additional 512 steps, after which the chains appear converged: \changestwo{We compute the Gelman-Rubin \citep{Gelman:1992} convergence diagnostic and find that all chains have $R < 1.1$.}
We then downsample the resulting chains by taking every 16th sample to preserve closer-to-independent samples, leaving a total of 1024 samples; We use these samples to estimate the median posterior parameter values and uncertainties,
For the young cluster, we find $\bs{x} = (-0.56,  0.47) \pm (0.04, 0.02)~\masyr$, $\ln s = -3.8 \pm 0.9$, and $f = 0.14 \pm 0.02$.
\changesthr{The proper-motion dispersion is consistent with zero, but is unmeasured: 95\% of the posterior samples have $s < 0.09~\masyr$, indicating that the observed proper motion dispersion is consistent with the proper motion \gaia\ uncertainties, which have minimum and median values of $\sim 0.09~\masyr$ and $\sim 0.4~\masyr$, respectively.}
\tablename~\ref{tbl:gaiadata} contains a subset of the \gaia\ \DR{2} data for stars in the cluster field, with kinematic membership probabilities determined from this modeling procedure.

\subsection{Inferring the stellar population parameters with \decam photometry}
\label{sec:popmodel}

While the \gaia\ data provide exquisite astrometric data for the brightest members of \clustername, constraints on the stellar population from the \gaia\ data alone are limited by degeneracies in the mapping from the color-magnitude diagram to stellar parameters.
\changes{In addition, uncertainty in the \gaia\ passbands can cause significant differences in \gaia\ \bprp\ color for very blue sources \citep{MAW:2018}.}
To measure the cluster population parameters, we therefore use the deeper photometry from \decam, which provides a clear view of the main sequence of \clustername\ down to stellar masses $M \sim 0.9~\msun$.
We note that this analysis and the subsequent discussion only considers subregion (a) of \clustername, which dominates the total mass and number of stars in the cluster.
\figurename~\ref{fig:decam-cmd} shows the \decam\ g- and i-band color-magnitude diagrams for control fields (left) and cluster fields (right) selected from the \decam\ footprint, once again showing the young main sequence (coherent stellar population in right panel).
We use the g- and i-band photometry---\changestwo{and u-band, though only the brightest $\sim$15 cluster field members are robustly detected, however for those sources, the u-band significantly helps to constrain the extinction for the cluster}---for individual stars in a sub-section of the cluster field CMD (red dashed outlined region in \figurename~\ref{fig:decam-cmd}) to infer the stellar population parameters of \clustername.
To summarize our methodology, we first generate independent posterior samplings over the stellar parameters of each individual source under an interim prior (over age, stellar mass, distance, etc.), then use these individual samplings to construct a Bayesian hierarchical model for a two-component mixture model of the cluster and a background population.
\changestwo{This methodology is more robust than conventional isochrone-fitting methods that require by-hand fitting of stellar population parameters and naturally propagates uncertainties in the photometry and cluster membership.}
If you prefer to skip the details, the results of this modeling are presented in \sectionname~\ref{sec:popchars}.

In detail, we start by using the \texttt{isochrones} package \citep{Morton:2015} to generate posterior samplings over stellar parameters for each individual source given its photometry and an interim prior.
We use the \acronym{MIST} \citep{Dotter:2016, Choi:2016, Paxton:2011, Paxton:2013, Paxton:2015} isochrone grid, and \texttt{isochrones} automatically performs interpolation between the provided grid of stellar isochrones to predict photometry given a set of stellar parameters.
Here, the stellar parameters for each source are its ``equal evolutionary point'' number $\eep$ (see \acronym{MIST} documentation\footnote{\url{http://waps.cfa.harvard.edu/MIST/}}), age $\tau$, metallicity $\feh$, extinction $A_V$, and distance $D$, which can be uniquely mapped to a point in the observed CMD;
The likelihood of a given set of these parameters, $\bs{\theta} = (\eep, \tau, \feh, A_V, D)$, is then computed from the photometry and (assumed Gaussian) photometric uncertainties given the predicted photometry. 

To generate posterior samples, we must also specify prior probability distributions over the parameters $\bs{\theta}$.
The priors for each parameter are summarized in \tablename~\ref{tbl:priors}.
The bounds on $\eep$ limit the isochrone to evolutionary phases between the zero-age main sequence to the terminal-age main sequence, but this prior is actually computed using the stellar mass computed from the isochrone parameters (see caption of \tablename~\ref{tbl:priors}).
The prior and bounds on $A_V$ are set to prefer small and reasonable values of extinction for this moderately high Galactic latitude region.
The prior on distance assumes a uniform space density of stars.

\begin{table}[htb]
\begin{center}
    \begin{tabular}{ c | c | c }
        \toprule
        Parameter & Prior & Bounds \\
        \toprule
        $\tau$ & uniform & $(10~\textrm{Myr}, 15~\textrm{Gyr})$ \\
        $\feh$ & uniform & $(-2, 0.5)$ \\
        $\eep$ & (see caption) & $(202, 355)$ \\
        $A_V$ & $\propto A_V^{-1}$ & $(0.001, 1)~\textrm{mag}$ \\
        $D$ & $\propto D^{2}$ & $(1, 100)~\kpc$ \\
        \toprule
    \end{tabular}
\caption{Prior probability distributions for each of the stellar parameters defined in \sectionname~\ref{sec:popmodel}, used for the independent (per star) posterior samplings.
The stellar parameters here are the ``equal evolutionary point'' number $\eep$, age $\tau$, metallicity $\feh$, extinction $A_V$, and distance $D$.
The prior on \eep\ is computed by calculating the stellar mass, $m$, corresponding to a given point in the isochrone parameter space, $(\eep, \tau, \feh)$, and computing the probability from a Salpeter \citep{Salpeter:1955} mass prior such that $p(\eep) = p(m)\,\left|\frac{\dd m}{\dd \eep}\right|$, following the prescription used in \texttt{isochrones} \citep{Morton:2015}.
\label{tbl:priors}}
\end{center}
\end{table}

We find that the photometry for individual sources in the lower main sequence are very poorly constrained in all parameters, and the prior tends to pull the posterior samplings to prefer closer, older stellar parameters.
This is a weakness of our methodology for performing the hierarchical inference: Each source is considered independently, even though there is clear structure in the CMD (i.e. \figurename~\ref{fig:decam-cmd}), and inferring isochronal parameters for individual lower main sequence stars is a fundamentally degenerate problem.
A more correct way to do this would be to infer the stellar parameters of \emph{all} stars, the cluster hyperparameters, and the background simultaneously.
However, for the \Nisofit\ stars we are using (in the red box in \figurename~\ref{fig:decam-cmd}), this model would have $\sim$2,000 free parameters if left unmarginalized.
We are developing tools to perform star cluster parameter inference in this way (Morton et al., in prep.), but here we adopt a simple hack to allow us to instead perform individual posterior samplings and then combine those samplings into a hierarchical inference.

The brightest stars in the cluster have very precise \decam\ photometry, and have precise astrometry from \gaia.
Given their location in the CMD, these stars must be young.
We therefore use the brightest \emph{kinematic} member of \clustername\ --- \gaia\ source ID 3480046557809199616 --- as an ``anchor'' star:
We model every other source in the \decam\ selection region (i.e. excluding this one) by fitting the photometry of it and the anchor star simultaneously, assuming they have the same age, metallicity, distance, and extinction but different \eep\ values.
Motivated by the possible signature of an (unresolved) binary sequence in the \decam\ CMD, we add one further piece of complexity to the model by allowing the photometry of each non-anchor source to be fit as an unresolved binary star system.
This adds an additional parameter, the unresolved binary mass ratio $q$, to the list of inferred stellar parameters for each source, but, for this work, we ignore the binary companions and implicitly marginalize over $q$ in what follows.
The details of this model are handled by the \texttt{isochrones} package.

We generate posterior samplings over the parameters $(\bs{\theta}, q)$ for each of the \Nisofit\ sources in the selected region of the \decam\ CMD using \texttt{PyMultinest} \citep{Buchner:2014, Feroz:2008, Feroz:2009}, and store the value of the prior evaluated at the location of each sample.
We then use these samples and interim prior values to construct our hierarchical model, as described below.

In the hierarchical model, we assume that the stellar parameters of each primary star in the selected region are either drawn from the cluster, or a background (stellar halo) population.
For the cluster, we assume that the values are drawn from delta functions in age, metallicity, distance, and extinction, with the centroids of the delta functions $\bs{\alpha} = (\tau^*, \feh^*, D^*, A_V^*)$ as hyperparameters of the hierarchical inference---\changesthr{that is, we assume that the cluster is a single stellar population (SSP).
We note that we have run the same hierarchical model allowing for Gaussian spreads in age, metallicity, and distance where we simultaneously infer the variances in each cluster parameter, but we have found that the variances are unconstrained, and we therefore instead treat the cluster as an SSP.}

For the background model, we assume the same priors as specified in \tablename~\ref{tbl:priors}.
The one additional parameter that must be included in this hierarchical model is the mixture weight: The global fraction of sources that are likely cluster members, $f$.
To compute the likelihood for the hierarchical model, we use the individual posterior samplings to marginalize over the per-source stellar parameters $\bs{\theta}_n$ to compute the marginal likelihood $p(\bs{m}_n \given \bs{\alpha}, f)$, where $\bs{m}_n = (g, i)_n$ is the vector of photometric data for source $n$;
This likelihood for a single source given a set of hyperparameters $(\bs{\alpha}, f)$ is then
\begin{equation}
    p(\bs{m}_n \given \bs{\alpha}, f) = \int \dd \bs{\theta}_n \,
        p(\bs{m}_n \given \bs{\theta}_n) \,
        p(\bs{\theta}_n \given \bs{\alpha}, f) \quad .
\end{equation}
We employ the ``importance sampling trick'' (see, e.g., Appendix of \citealt{Price-Whelan:2018} or \citealt{Hogg:2010, Foreman-Mackey:2014} for other examples) to re-write an approximate form for this marginal likelihood as
\begin{equation}
    p(\bs{m}_n \given \bs{\alpha}, f) \approx \frac{\mathcal{Z}_n}{K} \,
        \sum_k^K \frac{p(\bs{\theta}_{nk} \given \bs{\alpha}, f)}{p(\bs{\theta}_{nk} \given \bs{\alpha}_0, f)}
        \label{eq:marglike}
\end{equation}
where the index $k$ specifies the index of one of $K$ posterior samples generated from the independent samplings (described above), $\mathcal{Z}_n$ is a constant, and the denominator, $p(\bs{\theta}_{nk} \given \bs{\alpha}_0, f)$, are the values of the interim prior used to do the independent samplings.
In this work, $N=\Nisofit$ and we adopt $K=2048$.

\begin{figure}
\centering
\includegraphics[width=0.48\textwidth]{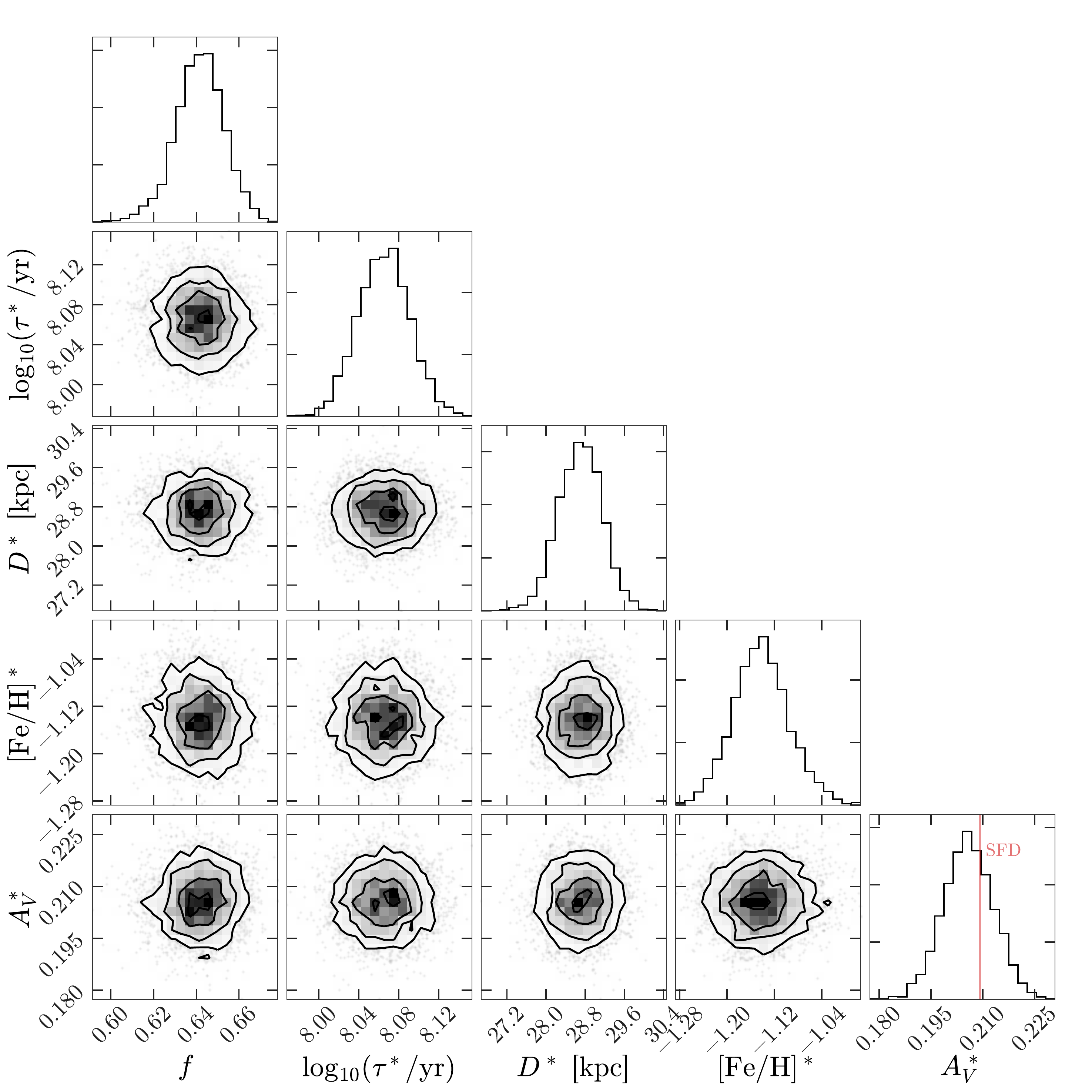}
\caption{A corner plot showing marginal posterior probability distributions estimated from the the posterior samples generated from the hierarchical inference of the cluster population parameters.}
\label{fig:hierarch-corner}
\end{figure}

With the marginal likelihood (\equationname~\ref{eq:marglike}), we then need to specify prior probability distributions for the hyperparameters $(\bs{\alpha}, f)$, and we can then generate posterior samples for the hyperparameters.
We use uniform priors for all of these, as summarized in \tablename~\ref{tbl:hyperpriors}.
We use \texttt{emcee} \citep{emcee, Goodman:2010} to sample from the posterior probability distribution for the hyperparameters given all of the photometric data,
\begin{equation}
    p(\bs{\alpha}, f \given \{\bs{m}_n\}) \propto
        p(\bs{\alpha}) \, p(f) \,
        \prod_n^N p(\bs{m}_n \given \bs{\alpha}, f) \quad.
\end{equation}
Here we use 64 walkers and run for an initial 128 steps to burn-in the sampler before running for a final 1024 steps.
\changestwo{We again compute the Gelman-Rubin \citep{Gelman:1992} convergence diagnostic and find that all chains have $R < 1.1$ and are thus likely converged.}
\figurename~\ref{fig:hierarch-corner} shows a corner plot with all 1D and 2D marginal posterior probability distributions estimated from the samples.

\begin{table}[hbt]
\begin{center}
    \begin{tabular}{ c | c | c }
        \toprule
        Parameter & Prior & Bounds \\
        \toprule
        $\tau^*$ & uniform & $(1~\textrm{Myr}, 1~\textrm{Gyr})$ \\
        $\feh^*$ & uniform & $(-2, 0)$ \\
        $A_V^*$ & uniform & $(0, 1)~\textrm{mag}$ \\
        $D^*$ & uniform & $(1, 100)~\kpc$ \\
        $f$ & uniform & (0, 1) \\
        \toprule
    \end{tabular}
\caption{Prior probability distributions for the hyperparameters $(\bs{\alpha}, f)$, used for the hierarchical inference of the cluster stellar population parameters.
The cluster parameters here are the cluster age $\tau^*$, metallicity $\feh^*$, extinction $A_V^*$, and distance $D^*$, and the fraction of stars in the field that belong to the cluster $f$.
\label{tbl:hyperpriors}}
\end{center}
\end{table}

\section{Results} \label{sec:results}

\subsection{Stellar population and physical characteristics}
\label{sec:popchars}

\begin{figure}[htb]
\centering
\includegraphics[width=0.45\textwidth]{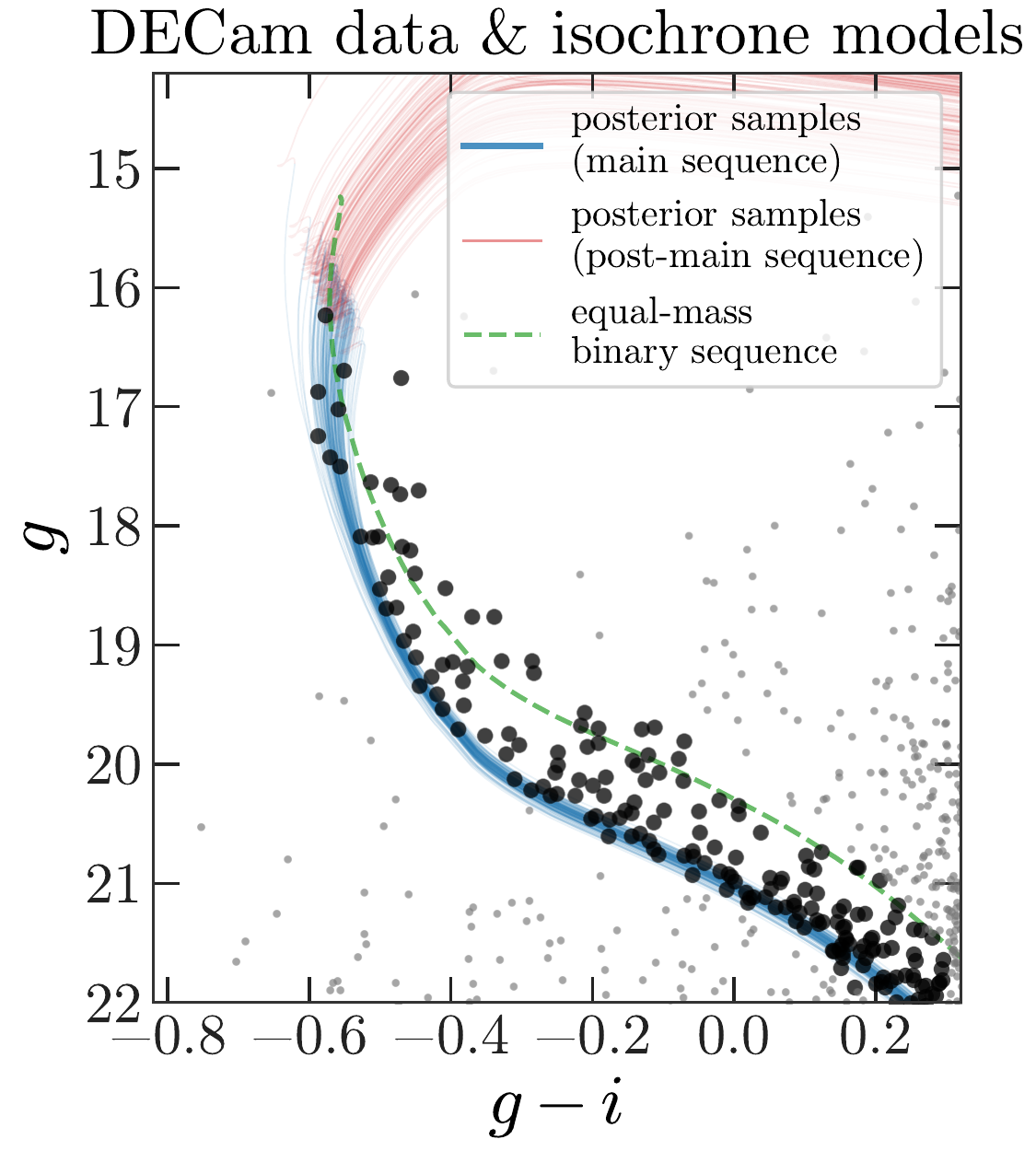}
\caption{The \decam\ color-magnitude diagram for sources with probability $>0.5$ of belonging to \clustername\ (dark, larger markers), and sources with probability $<0.5$ (lighter, smaller markers).
The solid line (blue) shows the \acronym{MIST} isochrone for the median posterior sample from the hierarchical inference of the cluster population parameters, shifted to the inferred distance to \clustername, and extincted with the inferred extinction.
The dashed line (green) shows the same isochrone, shifted $\approx 0.75$ magnitudes brighter, representing the expected location of the equal-mass binary sequence of the cluster.
}
\label{fig:hierarch-iso}
\end{figure}

We use the posterior samples from the hierarchical inference described in \sectionname~\ref{sec:popmodel} to compute the posterior probabilities that each source in the \decam\ CMD selection box (red box, \figurename~\ref{fig:decam-cmd}) is a member of the cluster.
\figurename~\ref{fig:hierarch-iso} shows the \decam\ photometry for all sources with membership probability $> 0.5$ (circle markers).
Over-plotted in \figurename~\ref{fig:hierarch-iso} (blue, solid line) is a \acronym{MIST} isochrone with the median posterior parameters derived from the hierarchical modeling, shifted to the median distance and extincted given the median $A_V$ value.
\changesthr{Under the assumption that \clustername\ has a single stellar population,} we find that \clustername\ is indeed young, distant, and metal poor, with median posterior values and standard deviations of age $\tau = \clageerr$, distance $D = \cldisterr$, and metallicity $\feh = \clfeherr$ (see also \tablename~\ref{tbl:clusterparams} for a summary).
We find that the inferred extinction, $A_V = \claverr$, is consistent with the value $A_{V, {\rm SFD}} \approx 0.2$ from the (recalibrated) SFD dust map \citep{Schlegel:1998, Schlafly:2011}.
We note again that, with the \decam\ photometry, we are limited to studying the subcomponent a of \clustername\ --- follow-up imaging of component b (see \figurename~\ref{fig:cmds}) would enable a similar analysis for the full cluster.

Also plotted in \figurename~\ref{fig:hierarch-iso} is the equal-mass binary sequence (green, dashed line) computed from the median posterior sample: The abundance of sources between the nominal isochrone and the binary sequence highlights the fact that the cluster may contain a significant number of binary or multiple star systems, but we leave a detailed study of multiplicity to future work.
\changes{We note that, similar to young open clusters in the Milky Way disk \citep[e.g.,][]{Babusiaux:2018}, the bluest stars in the CMD have more apparent scatter than the lower main sequence.
This is likely a combination of many things that are not addressed in this work, such as binarity, convective core overshooting \citep[e.g.,][]{Yang:2017, Johnston:2019}, or an intrinsic spread in stellar parameters.}

We use the isochrone corresponding to the median posterior sample to estimate the total stellar mass of the cluster.
By assuming that the \decam\ imaging is 100\% complete to stars with $(g-i) < 0.3)$ and $g < 22$, and by assuming a Kroupa initial mass function \citep{Kroupa:2001}, we use the number of observed stars and the isochrone to compute the total mass, $M_{\rm tot, *} \approx 1200~\msun$.

The mass and age of \clustername\ are comparable to Milky Way disk open clusters, but with a much lower metallicity, and a much larger spatial extent.
For example, the Pleiades has an age $\sim 135~\textrm{Myr}$ \citep{Gossage:2018}, but a physical size $\sim 5~\textrm{pc}$.
At a distance of $\cldist$, \clustername\ spans $\sim 1.5^\circ$, corresponding to a physical size $\sim 700$--$800~\textrm{pc}$.
This is more comparable to (but still larger than) recent star formation sites in the Magellanic bridge \citep[e.g.,][]{Mackey:2017}, which likely formed as a result of the violent interaction between the MCs.
If \clustername\ formed unbound, but with an initial size comparable to the present size of the Pleiades, this corresponds to an expansion velocity $\sim 6~\kms$, which would be detectable with precise radial velocity measurements of stars on either side of \clustername.

\begin{table}[htb]
\begin{center}
    \begin{tabular}{ r | l | l}
        \multicolumn{3}{c}{\textbf{Inferred properties of \clustername}} \\
        \toprule
        Name & Value & Description \\
        \tableline
        $\mean{\alpha}$ & $178.8^\circ$ & right ascension \\
        $\mean{\delta}$ & $-29.4^\circ$ & declination \\
        $\mean{D}$ & $28.7 \pm 0.4~\kpc$ & Heliocentric distance \\
        $\mean{\mu_\alpha}$ & $-0.56 \pm 0.04~\masyr$ & proper motion in RA\\
        $\mean{\mu_\delta}$ & $0.47 \pm 0.02~\masyr$ & proper motion in Dec\\
        \tableline
        $\tau$ & $116 \pm 7~\textrm{Myr}$ & age \\
        $M$ & $1200~\msun$ & total stellar mass \\
        $\feh$ & $-1.14 \pm 0.05$ & metallicity \\
        \tableline
        $A_V$ & $0.21 \pm 0.02$ & extinction \\
        \toprule
    \end{tabular}
\caption{This table summarizes the measured or inferred kinematic and stellar population parameters of the young halo stellar cluster \clustername.
Formal precisions on all inferred parameters are very small, typically one to a few per cent, but we suspect that systematic errors with the photometry and isochrone models limit the accuracy of these measurements to $\sim 5\%$.
\label{tbl:clusterparams}}
\end{center}
\end{table}

\subsection{Relation to the Magellanic stream}
\label{sec:higas}

\begin{figure}[t]
\centering
\includegraphics[width=8cm]{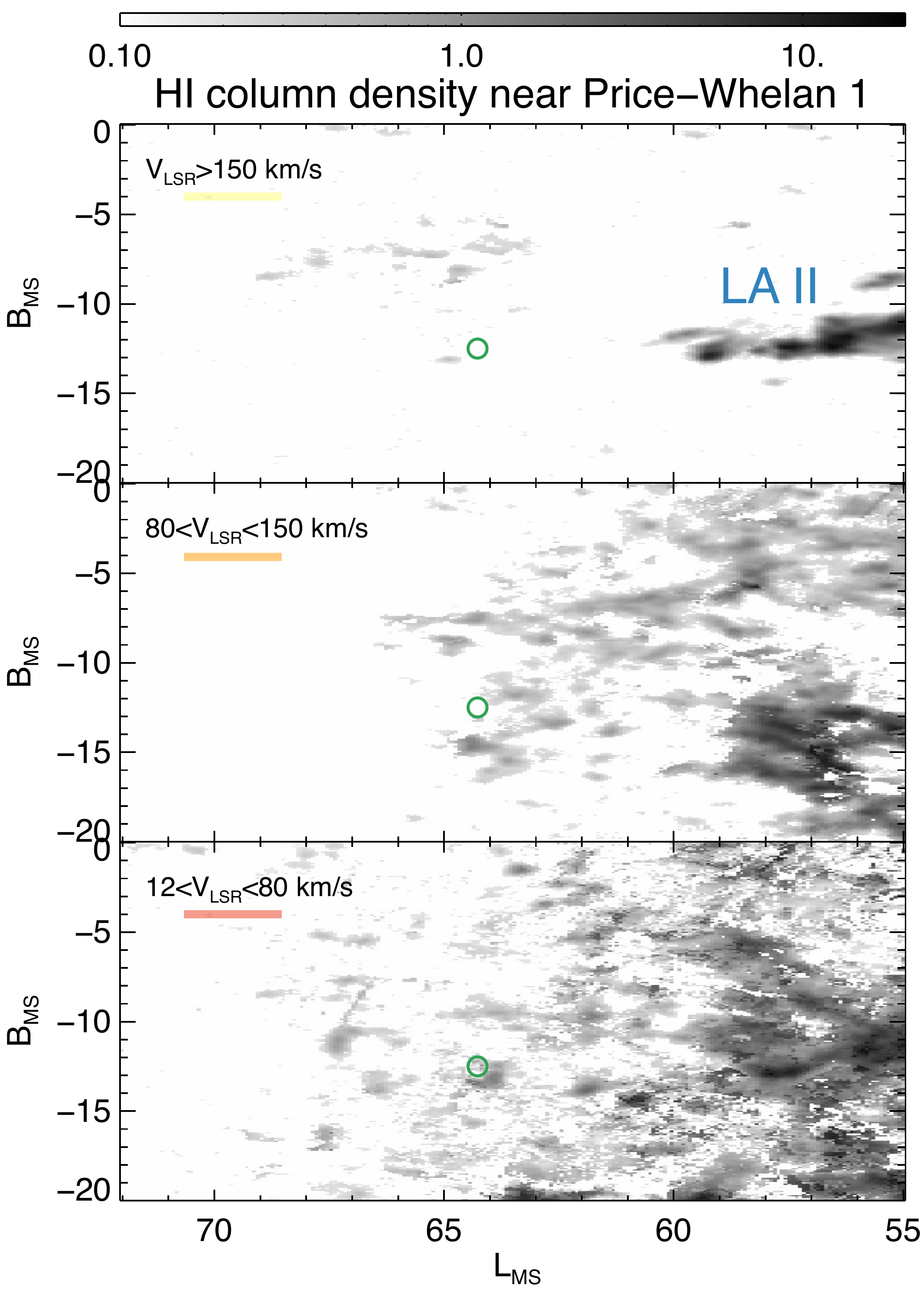}
\caption{GASS \hi column density in the region of LA II, shown in three different velocity slices (corresponding to and indicated in each panel).
The colored rectangles under each velocity label correspond to the colored ranges in \figurename~\ref{fig:gasspv}.
The coordinate system is the Magellanic stream spherical coordinate system defined in \cite{Nidever:2008}, and the units of $L_{\rm MS}$ and $B_{\rm MS}$ are degrees.
The horizontal colorbar shows the column density of \hi in units of $10^{19}~\textrm{atoms}~\textrm{cm}^{-2}$.
The open circle (green) marks the position of \clustername, and the feature commonly attributed to the LA is indicated in the top panel.}
\label{fig:gass_maps}
\end{figure}

\begin{figure}[t]
\centering
\includegraphics[width=8cm]{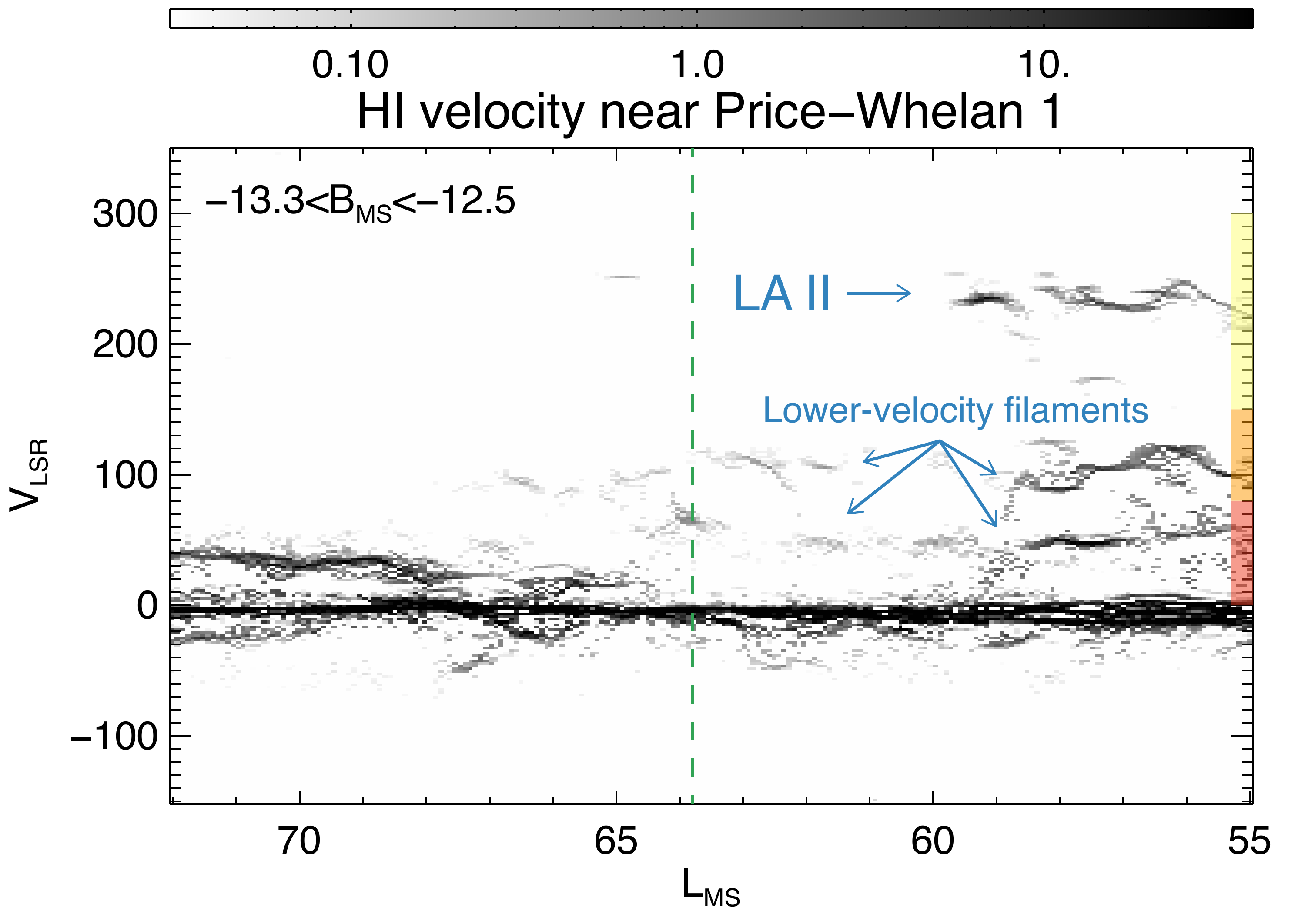}
\caption{The position-velocity diagram of GASS \hi Gaussian centers in a slice of Magellanic latitudes (indicated) in the same region as \figurename~\ref{fig:gass_maps}, showing the line-of-sight velocity (in the local standard of rest frame) of gas near \clustername.
The colored rectangles (red, orange, yellow) indicate the three velocity slices used in \figurename~\ref{fig:gass_maps}.
Here the colorbar shows the gas density in units of $10^{18}~\textrm{atoms}~\textrm{cm}^{-2}\,\textrm{deg}$.
Note that the densest gas at the longitude of \clustername\ (vertical line) appears to be associated with the lower velocity filaments seen at lower longitudes $L_{\textrm{MS}} \lesssim 60^\circ$, which may be decelerated gas associated with the Magellanic stream.
}
\label{fig:gasspv}
\end{figure}

At the sky location and distance of \clustername, i.e. well into the Galactic halo, the only plausible gas reservoirs that could have formed a young cluster are the MS, or a previously unknown high velocity cloud (HVC).
HVCs are thought to either be accreted and therefore lower metallicity than typical present-day Milky Way gas, or ejected from the Milky Way through a ``Galactic fountain''-like process and therefore comparable metallicity to disk gas.
Both processes clearly occur: the MS itself is evidence of gas accretion into the Milky Way halo, and the mysterious Smith Cloud \citep{Smith:1963} is a metal-rich \citep[$\feh \sim 0.5$][]{Fox:2016} HVC that plausibly originated from the Galactic disk \citep[e.g.,][]{Bregman:1980}.
Given the low metallicity of the stars in \clustername, the gas it formed from was likely extragalactic, as any violent star-forming regions in the Milky Way disk that could have driven gas so far out into the halo have, at present-day, significantly higher metallicity than this cluster.

We therefore take a closer look at the \hi gas in the vicinity of \clustername\ to assess its possible relation to the MS.
While no bulk metallicity measurements exist for the LA gas, recent measurements of oxygen and silicon abundances ($[\textrm{O}/\textrm{H}] \sim -1$) suggest that the gas has a comparable abundance pattern to the SMC gas \citep{Fox:2013, Fox:2018, Richter:2018}.
If \clustername\ formed from this gas, we would then expect that the stellar metallicity should be similar to young stars in the SMC.
In the outskirts of the SMC, the metallicity distribution of the youngest stars ($\textrm{age} < 2~\textrm{Gyr}$) has a peak around $\feh \sim -1$ and a spread of $\sigma_{\feh} \sim 0.2~\textrm{dex}$, which at least means that the metallicity of \clustername\ ($\feh \approx -1.1$) is consistent with young SMC stars \citep{Dobbie:2014}.

We next look at the kinematics of high velocity gas in this region.
\figurename~\ref{fig:gass_maps} shows the \hi column density of the catalog of Gaussian centers \citep[producted with the techniques and software from][]{Nidever:2008} from the Galactic All Sky Survey \citep[GASS;][]{McClure-Griffiths:2009, Kalberla:2010} in the region around the LA and plotted in the MS coordinate system \citep[$L_{\rm MS}, B_{\rm MS}$;][]{Nidever:2008}.
The location of \clustername\ is marked (green circle), and the three panels show three different velocity slices (indicated on each panel).
The MS gas is most visible and densest at the highest velocities where the LA II feature is found, around $L_{\rm MS} \lesssim 61^\circ$ and $B_{\rm MS} \sim -12^\circ$, but the lower panels (i.e. slower gas) still displays some of the same morphology as the highest velocity slice.

This is also apparent in a position-velocity diagram of the same range of longitudes:
\figurename~\ref{fig:gasspv} shows the velocity of gas in this region (all longitudes, but near $B_{\rm MS} \sim -13^\circ$) as a function of $L_{\rm MS}$.
Here, the longitude of \clustername\ is marked as the dashed vertical (green) line.
The LA II feature is again visible as the $V_{\textrm{LSR}} > 200~\kms$ filament of gas at $L_{\rm MS} \lesssim 60^\circ$ (and a faint extension near $L_{\rm MS} \sim 65^\circ$), but two other prominent filaments with similar morphologies are apparent at lower velocities $V_{\textrm{LSR}} \sim 60~\kms$ and $V_{\textrm{LSR}} \sim 100~\kms$ from $L_{\rm MS} \lesssim 65^\circ$.

\begin{figure*}[t!]
\centering
\includegraphics[width=0.9\textwidth]{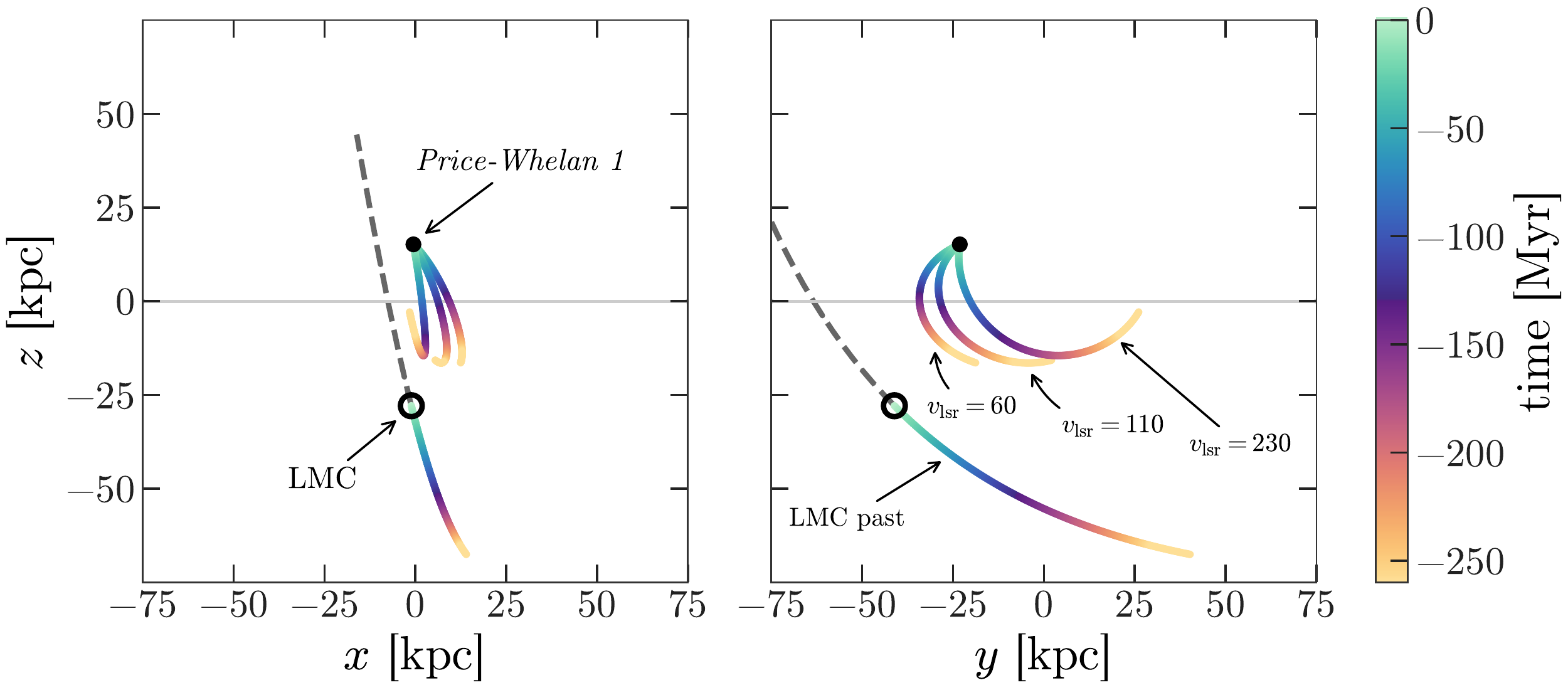}
\caption{The past orbit of \clustername in Galactocentric Cartesian coordinates, computed by assuming a line-of-sight velocity equal to each of three different possible gas  filaments in the LA II region of the MS (as indicated in right panel).
In these coordinates, the Sun is at $(x, y, z) = (-8.1, 0, 0)~\kpc$, and the present-day locations of \clustername\ (filled black marker) and the LMC (open black marker) are indicated.
The time at each point in the orbits is indicated by the colorbar, with the darkest regions corresponding to $t = -130~\textrm{Myr}$, the inferred age of \clustername.
The past and future orbit of the LMC are shown as the colored line and dashed line, respectively.
}
\label{fig:orbits}
\end{figure*}

The entire region shown in \figurename~\ref{fig:gass_maps} is on the other side of the Galactic disk with respect to the Magellanic clouds, meaning that all of the MS gas in this region has passed through the Galactic midplane.
The similar filamentary structure in the high and lower velocity gas therefore raises the question: Did the low velocity filaments originate in the MS, but were decelerated through interactions with Milky Way gas?
\changestwo{This scenario has been suggested (i.e., that the LA material has interacted with gas in the Galactic disk) based on the kinematics and morphology of a LA HVC \citep{McClure-Griffiths:2008}} and presents a possible formation scenario for \clustername.
If the cluster formed from the LA gas, and some of the gas interacted with the Milky Way disk, the dynamics of the stars and the gas that interacted would have decoupled as soon as the stars formed.
Given the geometry of the current location of \clustername\ (i.e. our viewing angle from the solar position), in this scenario we would expect a moderate velocity difference between the gas that interacted and stars to primarily appear as a difference in the tangential velocities of the stars and the gas (which is not measured).
We can therefore use the line-of-sight velocity of the gas as a proxy for the line-of-sight velocity of the stars in \clustername\ to study the Galactic orbit of the cluster, as discussed in the next section, and because we do not yet have radial velocity information for the stars.

\subsection{The Galactic orbit of \clustername}
\label{sec:orbit}

Since the gas distribution in \figurename~\ref{fig:gasspv} shows excess at velocities $V_{\textrm{LSR}} \approx (60, 110, 230)~\kms$ near the location of the cluster (ignoring the disk gas between $V_{\textrm{LSR}} \sim -20$--$20~\kms$), it seems that there are three qualitatively different line-of-sight velocities that the gas (and plausibly \clustername) could have.
If we adopt these velocity measurements as the line-of-sight (LOS) velocity of \clustername, we have measurements of all 6 phase-space components and can compute orbits for the cluster.
We compute Galactic orbits of the cluster using each of the three values of the line-of-sight velocity from the LA II region filaments.
\figurename~\ref{fig:orbits} shows these three orbits computed by integrating the position of \clustername\ backwards in time:
The two panels of \figurename~\ref{fig:orbits} show the Galactocentric Cartesian trajectories of \clustername\ over the last $260~\textrm{Myr}$ by assuming each of the three plausible line-of-sight velocities discussed above (as labeled in the right panel).
To compute the orbits, we use a three-component Milky Way mass model consisting of a disk \citep{Miyamoto:1975}, a bulge \citep{Hernquist:1990}, and a spherical dark matter halo \citep{Navarro:1996}.
The parameters of this model are set to match the circular velocity profile and disk properties of \citet{Bovy:2015}, and is implemented in the \texttt{gala} package \citep{gala} as the \texttt{MilkyWayPotential}.

We tried including an LMC component as a Hernquist sphere \citep{Hernquist:1990} with masses between $10^{10}$--$2.5\times 10^{11}~\msun$, but find that the orbit of \clustername\ does not change much over the integration period.
If the LMC mass were as high as $8\times 10^{11}~\msun$, all three orbits of \clustername\ return to and end up bound to the LMC, but at this large of a mass, the assumption of a fixed Milky Way reference frame is invalid.
We therefore neglect \lmcsmc\ component in what follows, and present an alternate interpretation below.

Also shown in \figurename~\ref{fig:orbits} is the present-day location of the LMC (open black circle), the present-day location of \clustername\ (filled black circle), the past orbit of the LMC over the same period, and the projected future orbit of the LMC (dashed line), all excluding the \lmcsmc\ as a mass component.
For the two lower LOS velocity cases, the orbit of \clustername\ crosses the midplane of the Galactic disk at a time comparable to the age of the cluster (i.e. when the color is darkest, corresponding to a time $\approx -130~\textrm{Myr}$).
However, while the orbits of \clustername\ and the LMC are generally in the same orbital plane and have the same sense about the Galaxy, none of the orbits closely approach (within the half-mass radius of our Hernquist model for the LMC) or cross the orbit of the LMC (even when including the LMC mass component).
This is expected if the gas was shocked or otherwise lost kinetic energy through, e.g., ram pressure --- as is expected to explain the morphology of the LA material \citep[e.g.,][]{Hammer:2015} --- during or before the star formation event that produced \clustername.
We therefore posit that the orbit of \clustername\ can be explained as having originated from the \lmcsmc\ system with some loss of orbital energy during the star formation event.

Precise radial velocity measurements of stars in \clustername\ will definitively rule on this scenario.
For one, with a precision $\sigma_v \lesssim 2~\kms$, the internal kinematics of \clustername\ could be resolved, and in particular the expansion rate could be measured.
But even low-precision velocity measurements would either confirm the association between \clustername\ and one of the three LA II-region gas filaments, or would challenge this scenario.

\subsection{Other possible formation scenarios}
\label{sec:otherformation}

While the spatial proximity of \clustername\ to LA gas in the Magellanic Stream suggests a tentative connection, radial velocity data for the stars is required to confirm that the velocity of the cluster is consistent with the gas and thereby confirm the connection (Nidever et al., in prep.).
However, even if the cluster can be confidently, kinematically associated with the LA gas, a number of mysteries would remain.
First, it is not clear what could have triggered this star formation event, or why this appears to be the only young cluster associated with the LA gas.
Second, the origin of the LA gas itself is not well understood, and some studies have questioned its connection to the Magellanic Clouds \citep{Tepper:2019}.

Other possible formation scenarios for \clustername\ do not help to resolve these issues.
One possible alternative scenario is that the \clustername\ birth cloud was unassociated with the MS and instead a lone HVC.
However, this still raises questions about what triggered the star formation, given that other HVCs in the Milky Way are observed to be devoid of stars (e.g., \citealt{Stark:2015}).
Another possible origin for \clustername\ is that it formed from gas stripped from a fully-destroyed dwarf galaxy (an unknown satellite of the Magellanic Clouds).
\changesthr{However, we have not found a significant over-density of comoving older stars associated with the LA in this region.}

\changesthr{One final possibility is that these stars were formed in or associated with the Galactic disk.
While there have been some claimed discoveries of young, embedded clusters at high Galactic latitude (within $\lesssim 5~\kpc$ of the sun, e.g., \citealt{Camargo:2015, Camargo:2016}, these may not be real \citep{Turner:2017}.
Still, stars formed in the Galactic disk have been observed far from the disk midplane with heights up to $|z| \sim 10~\kpc$ \citep[e.g., the TriAnd over-density;][]{Price-Whelan:2015, Bergemann:2018}).
However, TriAnd has an old stellar population, and no young disk stars have been since found associated with this or other features in the outer Galactic disk \citep{Deason:2018, Laporte:2019}.}

\section{Summary} \label{sec:conclusion}

We have identified a young, metal-poor stellar association in the Galactic halo --- named \clustername\ --- with an age $\tau \approx \clage$, heliocentric distance $D \approx \cldist$, and metallicity $\feh \approx \clfeh$.
At its present-day sky position, and at large distances, all significant quantities of \hi are associated with the leading arm of the Magellanic stream, and thus it plausibly formed from this gas.
The age of the cluster is broadly consistent with the time it would have most recently crossed the Galactic midplane, suggesting the possibility that interaction with the Milky Way disk or tidal compression could have triggered this star formation event.
Previous studies have detected young stars in the leading arm and in the periphery of the LMC \citep{Casetti-Dinescu:2014, Moni-Bidin:2017}, however, this is the first time that an entire young star cluster has been detected so far from the Clouds.

The discovery of \clustername\ provides a critical distance constraint to the leading arm Magellanic stream and will aid future Magellanic system and Milky Way modeling efforts.
It also provides an opportunity to study star formation in a unique environment (i.e. low gas density, low metallicity, and large velocity), unlike that of the Milky Way disk or any other local cluster-forming region.
The serendipitous discovery of this cluster is a reminder that the combined value of the \gaia\ data with deep, large-area imaging surveys provides a wealth of information about our Galaxy and stellar halo.

\acknowledgments

We thank Doug Finkbeiner (Harvard) for contributing to the \decam\ observations that enabled this work.

It is a pleasure to thank
Lauren Anderson (Flatiron),
Ana Bonaca (Harvard),
Elena D'Onghia (UW Madison),
Dan Foreman-Mackey (Flatiron),
Raja Guhathakurta (UCSC),
Cliff Johnson (Northwestern),
Semyeong Oh (Cambridge),
Ekta Patel (Arizona),
Sarah Pearson (Flatiron),
Josh Peek (STScI),
Anil Seth (Utah),
and Erik Tollerud (STScI)
for useful suggestions and discussion.

D.L.N thanks Peter Kalberla for making the GASS datacubes available in Magellanic Stream coordinates.

SK is supported by National Science Foundation grant AST-1813881.

The DECam results are based on observations at Cerro Tololo Inter-American Observatory, National Optical Astronomy Observatory (NOAO Prop. ID: 2018A-0251; PI: Finkbeiner), which is operated by the Association of Universities for Research in Astronomy (AURA) under a cooperative agreement with the National Science Foundation.

This work has made use of data from the European Space Agency (ESA)
mission {\it Gaia} (\url{https://www.cosmos.esa.int/gaia}), processed by
the {\it Gaia} Data Processing and Analysis Consortium (DPAC,
\url{https://www.cosmos.esa.int/web/gaia/dpac/consortium}). Funding
for the DPAC has been provided by national institutions, in particular
the institutions participating in the {\it Gaia} Multilateral Agreement.

We thank the Scientific Computing Core at the Flatiron Institute, and especially Dylan Simon and Nick Carriero, for technical support and access to the Flatiron Institute cluster computing resources, which enabled this work.
We thank the Center for Computational Astrophysics and especially David Spergel for support, access to computational resources, and space to conduct this work.

\clearpage

\appendix

\section{Queries}
\label{sec:queries}

Initial query to select very blue stars away from the Galactic plane:
\begin{verbatim}
SELECT * FROM gaiadr2.gaia_source
WHERE parallax < 1
AND (bp_rp > -0.5) AND (bp_rp < 0)
AND phot_g_mean_mag < 20
AND ABS(b) > 20
\end{verbatim}

Query to retrieve \gaia\ data around the blue, comoving group found and discussed in \sectionname~\ref{sec:data}:
\begin{verbatim}
SELECT *
FROM gaiadr2.gaia_source
WHERE (parallax < 1 OR parallax IS NULL)
    AND ra > 177 AND ra < 182
    AND dec > -31.3 AND dec < -26.3
\end{verbatim}

\section{Data}
\label{sec:datatables}

\begin{table}[h!]
    \centering
    \begin{tabular}{c | c | c | c | c | c | c | c | c}
        \multicolumn{9}{c}{\textbf{Catalog of \gaia\ sources and kinematic probability}}\\
        \texttt{source\_id} & \texttt{ra} & \texttt{dec} & \texttt{pmra} & \texttt{pmdec} & \texttt{G0} & \texttt{BP0} & \texttt{RP0} & \texttt{member\_prob} \\
        \tableline
        3462048793708019200 & 184.799 & -34.971 &  0.09 & -1.61 & 19.79 & 19.915 & 19.61 & 0.00 \\
        3462056696448285440 & 184.511 & -34.988 & -4.24 & -1.32 & 20.80 & 20.498 & 20.31 & 0.09 \\
        3462062670747949312 & 184.426 & -34.816 & -0.56 & -2.77 & 20.33 & 20.363 & 20.06 & 0.00 \\
        3462062709402484224 & 184.407 & -34.809 & -7.50 & -3.16 & 20.67 & 20.437 & 20.09 & 0.01 \\
        3462065904858152832 & 184.705 & -34.783 &  0.91 & -1.95 & 20.64 & 20.625 & 20.37 & 0.12 \\
        \multicolumn{9}{c}{...}\\
        \multicolumn{9}{c}{\textit{(5983 rows)}}
    \end{tabular}
    \caption{Select \gaia\ \DR{2} data and membership probabilities computed for all stars in the cluster region (see \figurename~\ref{fig:pm-members}).}
    \label{tbl:gaiadata}
\end{table}

\begin{table}[h!]
    \centering
    \begin{tabular}{c | c | c | c | c | c | c | c | c | c}
        \multicolumn{10}{c}{\textbf{Catalog of \decam\ photometry for point sources}}\\
        \texttt{name} & \texttt{ra} & \texttt{dec} & \texttt{g} & \texttt{g\_err} & \texttt{i} & \texttt{i\_err} & \texttt{control\_mask} & \texttt{cluster\_mask} & \texttt{member\_mask} \\
        \tableline
        J115311.06-283133.4 & 178.296 & -28.525 & 24.09 & 0.46 & 23.99 & 0.33 & True & False & False \\
        J115403.33-283553.8 & 178.513 & -28.598 & 25.27 & 0.29 & 22.70 & 0.04 & True & False & False \\
        J115424.21-283749.6 & 178.600 & -28.630 & 24.61 & 0.25 & 23.83 & 0.28 & True & False & False \\
        J115413.93-283343.6 & 178.558 & -28.562 & 25.37 & 0.41 & 22.76 & 0.05 & True & False & False \\
        J115419.62-283635.8 & 178.581 & -28.609 & 24.20 & 0.13 & 21.89 & 0.02 & True & False & False \\
        \multicolumn{10}{c}{...}\\
        \multicolumn{10}{c}{\textit{(51932 rows)}}
    \end{tabular}
    \caption{\decam\ photometry and sky positions for point sources in the observed \decam\ field.
    The boolean columns \texttt{control\_mask} and \texttt{cluster\_mask} are True when a given source is in the control or cluster CCDs, respectively (see \figurename~\ref{fig:decam-field}).
    The boolean column \texttt{member\_mask} is True when the source has a photometric membership probability $>0.5$ of belonging to the cluster stellar population (see \sectionname~\ref{sec:popmodel}).
    }
    \label{tbl:decamdata}
\end{table}

\facility{Blanco (DECam)}

\software{
    \package{Astropy} \citep{astropy, astropy:2018},
    \package{dustmaps} \citep{dustmaps},
    \package{emcee} \citep{emcee, emcee:ascl},
    \package{gala} \citep{gala},
    \package{IPython} \citep{ipython},
    \package{isochrones} \citep{Morton:2015},
    \package{matplotlib} \citep{mpl},
    \package{numpy} \citep{numpy},
    \package{PyMultinest} \citep{Buchner:2014},
    \package{schwimmbad} \citep{schwimmbad},
    \package{scipy} \citep{scipy}
}

\bibliographystyle{aasjournal}
\bibliography{ms}

\begin{thebibliography}{}
\expandafter\ifx\csname natexlab\endcsname\relax\def\natexlab#1{#1}\fi
\providecommand{\url}[1]{\href{#1}{#1}}
\providecommand{\dodoi}[1]{doi:~\href{http://doi.org/#1}{\nolinkurl{#1}}}
\providecommand{\doeprint}[1]{\href{http://ascl.net/#1}{\nolinkurl{http://ascl.net/#1}}}
\providecommand{\doarXiv}[1]{\href{https://arxiv.org/abs/#1}{\nolinkurl{https://arxiv.org/abs/#1}}}

\bibitem[{{Astropy Collaboration} {et~al.}(2013){Astropy Collaboration},
  {Robitaille}, {Tollerud}, {Greenfield}, {Droettboom}, {Bray}, {Aldcroft},
  {Davis}, {Ginsburg}, {Price-Whelan}, {Kerzendorf}, {Conley}, {Crighton},
  {Barbary}, {Muna}, {Ferguson}, {Grollier}, {Parikh}, {Nair}, {Unther},
  {Deil}, {Woillez}, {Conseil}, {Kramer}, {Turner}, {Singer}, {Fox}, {Weaver},
  {Zabalza}, {Edwards}, {Azalee Bostroem}, {Burke}, {Casey}, {Crawford},
  {Dencheva}, {Ely}, {Jenness}, {Labrie}, {Lim}, {Pierfederici}, {Pontzen},
  {Ptak}, {Refsdal}, {Servillat}, \& {Streicher}}]{astropy}
{Astropy Collaboration}, {Robitaille}, T.~P., {Tollerud}, E.~J., {et~al.} 2013,
  \aap, 558, A33, \dodoi{10.1051/0004-6361/201322068}

\bibitem[{{Astropy Collaboration} {et~al.}(2018){Astropy Collaboration},
  {Price-Whelan}, {Sip{\'{o}}cz}, {G{\"u}nther}, {Lim}, {Crawford}, {Conseil},
  {Shupe}, {Craig}, {Dencheva}, {Ginsburg}, {VanderPlas}, {Bradley},
  {P{\'e}rez-Su{\'a}rez}, {de Val- Borro}, {Aldcroft}, {Cruz}, {Robitaille},
  {Tollerud}, {Ardelean}, {Babej}, {Bach}, {Bachetti}, {Bakanov}, {Bamford},
  {Barentsen}, {Barmby}, {Baumbach}, {Berry}, {Biscani}, {Boquien}, {Bostroem},
  {Bouma}, {Brammer}, {Bray}, {Breytenbach}, {Buddelmeijer}, {Burke},
  {Calderone}, {Cano Rodr{\'\i}guez}, {Cara}, {Cardoso}, {Cheedella}, {Copin},
  {Corrales}, {Crichton}, {D'Avella}, {Deil}, {Depagne}, {Dietrich}, {Donath},
  {Droettboom}, {Earl}, {Erben}, {Fabbro}, {Ferreira}, {Finethy}, {Fox},
  {Garrison}, {Gibbons}, {Goldstein}, {Gommers}, {Greco}, {Greenfield},
  {Groener}, {Grollier}, {Hagen}, {Hirst}, {Homeier}, {Horton}, {Hosseinzadeh},
  {Hu}, {Hunkeler}, {Ivezi{\'c}}, {Jain}, {Jenness}, {Kanarek}, {Kendrew},
  {Kern}, {Kerzendorf}, {Khvalko}, {King}, {Kirkby}, {Kulkarni}, {Kumar},
  {Lee}, {Lenz}, {Littlefair}, {Ma}, {Macleod}, {Mastropietro}, {McCully},
  {Montagnac}, {Morris}, {Mueller}, {Mumford}, {Muna}, {Murphy}, {Nelson},
  {Nguyen}, {Ninan}, {N{\"o}the}, {Ogaz}, {Oh}, {Parejko}, {Parley}, {Pascual},
  {Patil}, {Patil}, {Plunkett}, {Prochaska}, {Rastogi}, {Reddy Janga},
  {Sabater}, {Sakurikar}, {Seifert}, {Sherbert}, {Sherwood-Taylor}, {Shih},
  {Sick}, {Silbiger}, {Singanamalla}, {Singer}, {Sladen}, {Sooley},
  {Sornarajah}, {Streicher}, {Teuben}, {Thomas}, {Tremblay}, {Turner},
  {Terr{\'o}n}, {van Kerkwijk}, {de la Vega}, {Watkins}, {Weaver}, {Whitmore},
  {Woillez}, {Zabalza}, \& {Astropy Contributors}}]{astropy:2018}
{Astropy Collaboration}, {Price-Whelan}, A.~M., {Sip{\'{o}}cz}, B.~M., {et~al.}
  2018, \aj, 156, 123, \dodoi{10.3847/1538-3881/aabc4f}

\bibitem[{{Bekki} {et~al.}(2008){Bekki}, {Chiba}, \&
  {McClure-Griffiths}}]{Bekki:2008}
{Bekki}, K., {Chiba}, M., \& {McClure-Griffiths}, N.~M. 2008, \apj, 672, L17,
  \dodoi{10.1086/526456}

\bibitem[{{Belokurov} \& {Erkal}(2019)}]{Belokurov:2019}
{Belokurov}, V.~A., \& {Erkal}, D. 2019, \mnras, 482, L9,
  \dodoi{10.1093/mnrasl/sly178}

\bibitem[{{Bergemann} {et~al.}(2018){Bergemann}, {Sesar}, {Cohen}, {Serenelli},
  {Sheffield}, {Li}, {Casagrande}, {Johnston}, {Laporte}, {Price-Whelan},
  {Sch{\"o}nrich}, \& {Gould}}]{Bergemann:2018}
{Bergemann}, M., {Sesar}, B., {Cohen}, J.~G., {et~al.} 2018, Nature, 555, 334,
  \dodoi{10.1038/nature25490}

\bibitem[{{Besla} {et~al.}(2007){Besla}, {Kallivayalil}, {Hernquist},
  {Robertson}, {Cox}, {van der Marel}, \& {Alcock}}]{Besla:2007}
{Besla}, G., {Kallivayalil}, N., {Hernquist}, L., {et~al.} 2007, \apj, 668,
  949, \dodoi{10.1086/521385}

\bibitem[{{Besla} {et~al.}(2012){Besla}, {Kallivayalil}, {Hernquist}, {van der
  Marel}, {Cox}, \& {Kere{\v s}}}]{Besla:2012}
---. 2012, \mnras, 421, 2109, \dodoi{10.1111/j.1365-2966.2012.20466.x}

\bibitem[{{Besla} {et~al.}(2010){Besla}, {Kallivayalil}, {Hernquist}, {van der
  Marel}, {Cox}, \& {Kere{\v{s}}}}]{Besla:2010}
---. 2010, \apj, 721, L97, \dodoi{10.1088/2041-8205/721/2/L97}

\bibitem[{{Bovy}(2015)}]{Bovy:2015}
{Bovy}, J. 2015, The Astrophysical Journal Supplement Series, 216,
  \dodoi{10.1088/0067-0049/216/2/29}

\bibitem[{Bovy {et~al.}(2011)Bovy, Hogg, \& Roweis}]{Bovy:2011}
Bovy, J., Hogg, D.~W., \& Roweis, S.~T. 2011, Ann. Appl. Stat., 5, 1657,
  \dodoi{10.1214/10-AOAS439}

\bibitem[{{Bregman}(1980)}]{Bregman:1980}
{Bregman}, J.~N. 1980, \apj, 236, 577, \dodoi{10.1086/157776}

\bibitem[{{Br{\"u}ns} {et~al.}(2005){Br{\"u}ns}, {Kerp}, {Staveley-Smith},
  {Mebold}, {Putman}, {Haynes}, {Kalberla}, {Muller}, \&
  {Filipovic}}]{Bruns:2005}
{Br{\"u}ns}, C., {Kerp}, J., {Staveley-Smith}, L., {et~al.} 2005, \aap, 432,
  45, \dodoi{10.1051/0004-6361:20040321}

\bibitem[{{Buchner} {et~al.}(2014){Buchner}, {Georgakakis}, {Nandra}, {Hsu},
  {Rangel}, {Brightman}, {Merloni}, {Salvato}, {Donley}, \&
  {Kocevski}}]{Buchner:2014}
{Buchner}, J., {Georgakakis}, A., {Nandra}, K., {et~al.} 2014, \aap, 564, A125,
  \dodoi{10.1051/0004-6361/201322971}

\bibitem[{{Burton} \& {Lockman}(1999)}]{Burton:1999}
{Burton}, W.~B., \& {Lockman}, F.~J. 1999, \aap, 349, 7.
\newblock \doarXiv{astro-ph/9908012}

\bibitem[{{Camargo} {et~al.}(2016){Camargo}, {Bica}, \&
  {Bonatto}}]{Camargo:2016}
{Camargo}, D., {Bica}, E., \& {Bonatto}, C. 2016, \aap, 593, A95,
  \dodoi{10.1051/0004-6361/201628710}

\bibitem[{{Camargo} {et~al.}(2015){Camargo}, {Bica}, {Bonatto}, \&
  {Salerno}}]{Camargo:2015}
{Camargo}, D., {Bica}, E., {Bonatto}, C., \& {Salerno}, G. 2015, \mnras, 448,
  1930, \dodoi{10.1093/mnras/stv092}

\bibitem[{{Carrera} {et~al.}(2017){Carrera}, {Conn}, {No{\"e}l}, {Read}, \&
  {L{\'o}pez S{\'a}nchez}}]{Carrera:2017}
{Carrera}, R., {Conn}, B.~C., {No{\"e}l}, N.~E.~D., {Read}, J.~I., \&
  {L{\'o}pez S{\'a}nchez}, {\'A}.~R. 2017, \mnras, 471, 4571,
  \dodoi{10.1093/mnras/stx1932}

\bibitem[{{Casetti-Dinescu} {et~al.}(2014){Casetti-Dinescu}, {Moni Bidin},
  {Girard}, {M{\'e}ndez}, {Vieira}, {Korchagin}, \& {van
  Altena}}]{Casetti-Dinescu:2014}
{Casetti-Dinescu}, D.~I., {Moni Bidin}, C., {Girard}, T.~M., {et~al.} 2014,
  \apjl, 784, L37, \dodoi{10.1088/2041-8205/784/2/L37}

\bibitem[{{Chambers} {et~al.}(2016){Chambers}, {Magnier}, {Metcalfe},
  {Flewelling}, {Huber}, {Waters}, {Denneau}, {Draper}, {Farrow}, {Finkbeiner},
  {Holmberg}, {Koppenhoefer}, {Price}, {Saglia}, {Schlafly}, {Smartt},
  {Sweeney}, {Wainscoat}, {Burgett}, {Grav}, {Heasley}, {Hodapp}, {Jedicke},
  {Kaiser}, {Kudritzki}, {Luppino}, {Lupton}, {Monet}, {Morgan}, {Onaka},
  {Stubbs}, {Tonry}, {Banados}, {Bell}, {Bender}, {Bernard}, {Botticella},
  {Casertano}, {Chastel}, {Chen}, {Chen}, {Cole}, {Deacon}, {Frenk},
  {Fitzsimmons}, {Gezari}, {Goessl}, {Goggia}, {Goldman}, {Grebel}, {Hambly},
  {Hasinger}, {Heavens}, {Heckman}, {Henderson}, {Henning}, {Holman}, {Hopp},
  {Ip}, {Isani}, {Keyes}, {Koekemoer}, {Kotak}, {Long}, {Lucey}, {Liu},
  {Martin}, {McLean}, {Morganson}, {Murphy}, {Nieto-Santisteban}, {Norberg},
  {Peacock}, {Pier}, {Postman}, {Primak}, {Rae}, {Rest}, {Riess}, {Riffeser},
  {Rix}, {Roser}, {Schilbach}, {Schultz}, {Scolnic}, {Szalay}, {Seitz},
  {Shiao}, {Small}, {Smith}, {Soderblom}, {Taylor}, {Thakar}, {Thiel},
  {Thilker}, {Urata}, {Valenti}, {Walter}, {Watters}, {Werner}, {White},
  {Wood-Vasey}, \& {Wyse}}]{Chambers:2016}
{Chambers}, K.~C., {Magnier}, E.~A., {Metcalfe}, N., {et~al.} 2016, ArXiv
  e-prints.
\newblock \doarXiv{1612.05560}

\bibitem[{{Choi} {et~al.}(2016){Choi}, {Dotter}, {Conroy}, {Cantiello},
  {Paxton}, \& {Johnson}}]{Choi:2016}
{Choi}, J., {Dotter}, A., {Conroy}, C., {et~al.} 2016, \apj, 823, 102,
  \dodoi{10.3847/0004-637X/823/2/102}

\bibitem[{{Choi} {et~al.}(2018{\natexlab{a}}){Choi}, {Nidever}, {Olsen},
  {Blum}, {Besla}, {Zaritsky}, {van der Marel}, {Bell}, {Gallart}, {Cioni},
  {Johnson}, {Vivas}, {Saha}, {de Boer}, {No{\"e}l}, {Monachesi}, {Massana},
  {Conn}, {Martinez-Delgado}, {Mu{\~n}oz}, \& {Stringfellow}}]{Choi:2018a}
{Choi}, Y., {Nidever}, D.~L., {Olsen}, K., {et~al.} 2018{\natexlab{a}}, \apj,
  866, 90, \dodoi{10.3847/1538-4357/aae083}

\bibitem[{{Choi} {et~al.}(2018{\natexlab{b}}){Choi}, {Nidever}, {Olsen},
  {Besla}, {Blum}, {Zaritsky}, {Cioni}, {van der Marel}, {Bell}, {Johnson},
  {Vivas}, {Walker}, {de Boer}, {Noel}, {Monachesi}, {Gallart}, {Monelli},
  {Stringfellow}, {Massana}, \& {Martinez-Delgado}}]{Choi:2018b}
---. 2018{\natexlab{b}}, ArXiv e-prints.
\newblock \doarXiv{1805.00481}

\bibitem[{{Danielski} {et~al.}(2018){Danielski}, {Babusiaux}, {Ruiz-Dern},
  {Sartoretti}, \& {Arenou}}]{Danielski:2018}
{Danielski}, C., {Babusiaux}, C., {Ruiz-Dern}, L., {Sartoretti}, P., \&
  {Arenou}, F. 2018, \aap, 614, A19, \dodoi{10.1051/0004-6361/201732327}

\bibitem[{{Deason} {et~al.}(2018){Deason}, {Belokurov}, \&
  {Koposov}}]{Deason:2018}
{Deason}, A.~J., {Belokurov}, V., \& {Koposov}, S.~E. 2018, Monthly Notices of
  the Royal Astronomical Society, 473, 2428, \dodoi{10.1093/mnras/stx2528}

\bibitem[{{Deason} {et~al.}(2015){Deason}, {Belokurov}, \&
  {Weisz}}]{Deason:2015}
{Deason}, A.~J., {Belokurov}, V., \& {Weisz}, D.~R. 2015, \mnras, 448, L77,
  \dodoi{10.1093/mnrasl/slv001}

\bibitem[{{Diaz} \& {Bekki}(2012)}]{Diaz:2012}
{Diaz}, J.~D., \& {Bekki}, K. 2012, \apj, 750, 36,
  \dodoi{10.1088/0004-637X/750/1/36}

\bibitem[{{Dobbie} {et~al.}(2014){Dobbie}, {Cole}, {Subramaniam}, \&
  {Keller}}]{Dobbie:2014}
{Dobbie}, P.~D., {Cole}, A.~A., {Subramaniam}, A., \& {Keller}, S. 2014,
  \mnras, 442, 1680, \dodoi{10.1093/mnras/stu926}

\bibitem[{{D'Onghia} \& {Fox}(2016)}]{DOnghia:2016}
{D'Onghia}, E., \& {Fox}, A.~J. 2016, Annual Review of Astronomy and
  Astrophysics, 54, 363, \dodoi{10.1146/annurev-astro-081915-023251}

\bibitem[{{Dotter}(2016)}]{Dotter:2016}
{Dotter}, A. 2016, \apjs, 222, 8, \dodoi{10.3847/0067-0049/222/1/8}

\bibitem[{{Feroz} \& {Hobson}(2008)}]{Feroz:2008}
{Feroz}, F., \& {Hobson}, M.~P. 2008, \mnras, 384, 449,
  \dodoi{10.1111/j.1365-2966.2007.12353.x}

\bibitem[{{Feroz} {et~al.}(2009){Feroz}, {Hobson}, \& {Bridges}}]{Feroz:2009}
{Feroz}, F., {Hobson}, M.~P., \& {Bridges}, M. 2009, \mnras, 398, 1601,
  \dodoi{10.1111/j.1365-2966.2009.14548.x}

\bibitem[{{Fiorentino} {et~al.}(2015){Fiorentino}, {Bono}, {Monelli},
  {Stetson}, {Tolstoy}, {Gallart}, {Salaris}, {Mart{\'\i}nez-V{\'a}zquez}, \&
  {Bernard}}]{Fiorentino:2015}
{Fiorentino}, G., {Bono}, G., {Monelli}, M., {et~al.} 2015, \apj, 798, L12,
  \dodoi{10.1088/2041-8205/798/1/L12}

\bibitem[{{For} {et~al.}(2014){For}, {Staveley-Smith}, {Matthews}, \& {McClure-
  Griffiths}}]{For:2014}
{For}, B.~Q., {Staveley-Smith}, L., {Matthews}, D., \& {McClure- Griffiths},
  N.~M. 2014, \apj, 792, 43, \dodoi{10.1088/0004-637X/792/1/43}

\bibitem[{{Foreman-Mackey} {et~al.}(2013{\natexlab{a}}){Foreman-Mackey},
  {Hogg}, {Lang}, \& {Goodman}}]{emcee}
{Foreman-Mackey}, D., {Hogg}, D.~W., {Lang}, D., \& {Goodman}, J.
  2013{\natexlab{a}}, Publications of the Astronomical Society of the Pacific,
  125, 306, \dodoi{10.1086/670067}

\bibitem[{{Foreman-Mackey} {et~al.}(2014){Foreman-Mackey}, {Hogg}, \&
  {Morton}}]{Foreman-Mackey:2014}
{Foreman-Mackey}, D., {Hogg}, D.~W., \& {Morton}, T.~D. 2014, \apj, 795, 64,
  \dodoi{10.1088/0004-637X/795/1/64}

\bibitem[{{Foreman-Mackey} {et~al.}(2013{\natexlab{b}}){Foreman-Mackey},
  {Conley}, {Meierjurgen Farr}, {Hogg}, {Lang}, {Marshall}, {Price-Whelan},
  {Sanders}, \& {Zuntz}}]{emcee:ascl}
{Foreman-Mackey}, D., {Conley}, A., {Meierjurgen Farr}, W., {et~al.}
  2013{\natexlab{b}}, {emcee: The MCMC Hammer}, Astrophysics Source Code
  Library.
\newblock \doeprint{1303.002}

\bibitem[{{Fox} {et~al.}(2013){Fox}, {Richter}, {Wakker}, {Lehner}, {Howk},
  {Ben Bekhti}, {Bland-Hawthorn}, \& {Lucas}}]{Fox:2013}
{Fox}, A.~J., {Richter}, P., {Wakker}, B.~P., {et~al.} 2013, \apj, 772, 110,
  \dodoi{10.1088/0004-637X/772/2/110}

\bibitem[{{Fox} {et~al.}(2016){Fox}, {Lehner}, {Lockman}, {Wakker}, {Hill},
  {Heitsch}, {Stark}, {Barger}, {Sembach}, \& {Rahman}}]{Fox:2016}
{Fox}, A.~J., {Lehner}, N., {Lockman}, F.~J., {et~al.} 2016, \apj, 816, L11,
  \dodoi{10.3847/2041-8205/816/1/L11}

\bibitem[{{Fox} {et~al.}(2018){Fox}, {Barger}, {Wakker}, {Richter},
  {Antwi-Danso}, {Casetti-Dinescu}, {Howk}, {Lehner}, {D'Onghia}, {Crowther},
  \& {Lockman}}]{Fox:2018}
{Fox}, A.~J., {Barger}, K.~A., {Wakker}, B.~P., {et~al.} 2018, \apj, 854, 142,
  \dodoi{10.3847/1538-4357/aaa9bb}

\bibitem[{{Gaia Collaboration} {et~al.}(2018{\natexlab{a}}){Gaia
  Collaboration}, {Brown}, {Vallenari}, {Prusti}, {de Bruijne}, {Babusiaux}, \&
  {Bailer-Jones}}]{Gaia-Collaboration:2018}
{Gaia Collaboration}, {Brown}, A.~G.~A., {Vallenari}, A., {et~al.}
  2018{\natexlab{a}}, ArXiv e-prints.
\newblock \doarXiv{1804.09365}

\bibitem[{{Gaia Collaboration} {et~al.}(2016){Gaia Collaboration}, {Prusti},
  {de Bruijne}, {Brown}, {Vallenari}, {Babusiaux}, {Bailer-Jones}, {Bastian},
  {Biermann}, {Evans}, {Eyer}, {Jansen}, {Jordi}, {Klioner}, {Lammers},
  {Lindegren}, {Luri}, {Mignard}, {Milligan}, {Panem}, {Poinsignon},
  {Pourbaix}, {Randich}, {Sarri}, {Sartoretti}, {Siddiqui}, {Soubiran},
  {Valette}, {van Leeuwen}, {Walton}, {Aerts}, {Arenou}, {Cropper}, {Drimmel},
  {H{\o}g}, {Katz}, {Lattanzi}, {O'Mullane}, {Grebel}, {Holland}, {Huc},
  {Passot}, {Bramante}, {Cacciari}, {Casta{\~n}eda}, {Chaoul}, {Cheek}, {De
  Angeli}, {Fabricius}, {Guerra}, {Hern{\'a}ndez}, {Jean-Antoine-Piccolo},
  {Masana}, {Messineo}, {Mowlavi}, {Nienartowicz}, {Ord{\'o}{\~n}ez- Blanco},
  {Panuzzo}, {Portell}, {Richards}, {Riello}, {Seabroke}, {Tanga},
  {Th{\'e}venin}, {Torra}, {Els}, {Gracia- Abril}, {Comoretto},
  {Garcia-Reinaldos}, {Lock}, {Mercier}, {Altmann}, {Andrae}, {Astraatmadja},
  {Bellas-Velidis}, {Benson}, {Berthier}, {Blomme}, {Busso}, {Carry},
  {Cellino}, {Clementini}, {Cowell}, {Creevey}, {Cuypers}, {Davidson}, {De
  Ridder}, {de Torres}, {Delchambre}, {Dell'Oro}, {Ducourant}, {Fr{\'e}mat},
  {Garc{\'\i}a-Torres}, {Gosset}, {Halbwachs}, {Hambly}, {Harrison}, {Hauser},
  {Hestroffer}, {Hodgkin}, {Huckle}, {Hutton}, {Jasniewicz}, {Jordan},
  {Kontizas}, {Korn}, {Lanzafame}, {Manteiga}, {Moitinho}, {Muinonen},
  {Osinde}, {Pancino}, {Pauwels}, {Petit}, {Recio-Blanco}, {Robin}, {Sarro},
  {Siopis}, {Smith}, {Smith}, {Sozzetti}, {Thuillot}, {van Reeven}, {Viala},
  {Abbas}, {Abreu Aramburu}, {Accart}, {Aguado}, {Allan}, {Allasia},
  {Altavilla}, {{\'A}lvarez}, {Alves}, {Anderson}, {Andrei}, {Anglada Varela},
  {Antiche}, {Antoja}, {Ant{\'o}n}, {Arcay}, {Atzei}, {Ayache}, {Bach},
  {Baker}, {Balaguer-N{\'u}{\~n}ez}, {Barache}, {Barata}, {Barbier}, {Barblan},
  {Baroni}, {Barrado y Navascu{\'e}s}, {Barros}, {Barstow}, {Becciani},
  {Bellazzini}, {Bellei}, {Bello Garc{\'\i}a}, {Belokurov}, {Bendjoya},
  {Berihuete}, {Bianchi}, {Bienaym{\'e}}, {Billebaud}, {Blagorodnova},
  {Blanco-Cuaresma}, {Boch}, {Bombrun}, {Borrachero}, {Bouquillon}, {Bourda},
  {Bouy}, {Bragaglia}, {Breddels}, {Brouillet}, {Br{\"u}semeister},
  {Bucciarelli}, {Budnik}, {Burgess}, {Burgon}, {Burlacu}, {Busonero}, {Buzzi},
  {Caffau}, {Cambras}, {Campbell}, {Cancelliere}, {Cantat-Gaudin}, {Carlucci},
  {Carrasco}, {Castellani}, {Charlot}, {Charnas}, {Charvet}, {Chassat},
  {Chiavassa}, {Clotet}, {Cocozza}, {Collins}, {Collins}, {Costigan}, {Crifo},
  {Cross}, {Crosta}, {Crowley}, {Dafonte}, {Damerdji}, {Dapergolas}, {David},
  {David}, {De Cat}, {de Felice}, {de Laverny}, {De Luise}, {De March}, {de
  Martino}, {de Souza}, {Debosscher}, {del Pozo}, {Delbo}, {Delgado},
  {Delgado}, {di Marco}, {Di Matteo}, {Diakite}, {Distefano}, {Dolding}, {Dos
  Anjos}, {Drazinos}, {Dur{\'a}n}, {Dzigan}, {Ecale}, {Edvardsson}, {Enke},
  {Erdmann}, {Escolar}, {Espina}, {Evans}, {Eynard Bontemps}, {Fabre},
  {Fabrizio}, {Faigler}, {Falc{\~a}o}, {Farr{\`a}s Casas}, {Faye}, {Federici},
  {Fedorets}, {Fern{\'a}ndez-Hern{\'a}ndez}, {Fernique}, {Fienga}, {Figueras},
  {Filippi}, {Findeisen}, {Fonti}, {Fouesneau}, {Fraile}, {Fraser}, {Fuchs},
  {Furnell}, {Gai}, {Galleti}, {Galluccio}, {Garabato}, {Garc{\'\i}a-Sedano},
  {Gar{\'e}}, {Garofalo}, {Garralda}, {Gavras}, {Gerssen}, {Geyer}, {Gilmore},
  {Girona}, {Giuffrida}, {Gomes}, {Gonz{\'a}lez-Marcos},
  {Gonz{\'a}lez-N{\'u}{\~n}ez}, {Gonz{\'a}lez-Vidal}, {Granvik}, {Guerrier},
  {Guillout}, {Guiraud}, {G{\'u}rpide}, {Guti{\'e}rrez-S{\'a}nchez}, {Guy},
  {Haigron}, {Hatzidimitriou}, {Haywood}, {Heiter}, {Helmi}, {Hobbs},
  {Hofmann}, {Holl}, {Holland}, {Hunt}, {Hypki}, {Icardi}, {Irwin}, {Jevardat
  de Fombelle}, {Jofr{\'e}}, {Jonker}, {Jorissen}, {Julbe}, {Karampelas},
  {Kochoska}, {Kohley}, {Kolenberg}, {Kontizas}, {Koposov}, {Kordopatis},
  {Koubsky}, {Kowalczyk}, {Krone-Martins}, {Kudryashova}, {Kull}, {Bachchan},
  {Lacoste-Seris}, {Lanza}, {Lavigne}, {Le Poncin-Lafitte}, {Lebreton},
  {Lebzelter}, {Leccia}, {Leclerc}, {Lecoeur-Taibi}, {Lemaitre}, {Lenhardt},
  {Leroux}, {Liao}, {Licata}, {Lindstr{\o}m}, {Lister}, {Livanou}, {Lobel},
  {L{\"o}ffler}, {L{\'o}pez}, {Lopez-Lozano}, {Lorenz}, {Loureiro},
  {MacDonald}, {Magalh{\~a}es Fernandes}, {Managau}, {Mann}, {Mantelet},
  {Marchal}, {Marchant}, {Marconi}, {Marie}, {Marinoni}, {Marrese},
  {Marschalk{\'o}}, {Marshall}, {Mart{\'\i}n-Fleitas}, {Martino}, {Mary},
  {Matijevi{\v{c}}}, {Mazeh}, {McMillan}, {Messina}, {Mestre}, {Michalik},
  {Millar}, {Miranda}, {Molina}, {Molinaro}, {Molinaro}, {Moln{\'a}r},
  {Moniez}, {Montegriffo}, {Monteiro}, {Mor}, {Mora}, {Morbidelli}, {Morel},
  {Morgenthaler}, {Morley}, {Morris}, {Mulone}, {Muraveva}, {Musella},
  {Narbonne}, {Nelemans}, {Nicastro}, {Noval}, {Ord{\'e}novic},
  {Ordieres-Mer{\'e}}, {Osborne}, {Pagani}, {Pagano}, {Pailler}, {Palacin},
  {Palaversa}, {Parsons}, {Paulsen}, {Pecoraro}, {Pedrosa}, {Pentik{\"a}inen},
  {Pereira}, {Pichon}, {Piersimoni}, {Pineau}, {Plachy}, {Plum}, {Poujoulet},
  {Pr{\v{s}}a}, {Pulone}, {Ragaini}, {Rago}, {Rambaux}, {Ramos-Lerate},
  {Ranalli}, {Rauw}, {Read}, {Regibo}, {Renk}, {Reyl{\'e}}, {Ribeiro},
  {Rimoldini}, {Ripepi}, {Riva}, {Rixon}, {Roelens}, {Romero-G{\'o}mez},
  {Rowell}, {Royer}, {Rudolph}, {Ruiz-Dern}, {Sadowski}, {Sagrist{\`a}
  Sell{\'e}s}, {Sahlmann}, {Salgado}, {Salguero}, {Sarasso}, {Savietto},
  {Schnorhk}, {Schultheis}, {Sciacca}, {Segol}, {Segovia}, {Segransan},
  {Serpell}, {Shih}, {Smareglia}, {Smart}, {Smith}, {Solano}, {Solitro},
  {Sordo}, {Soria Nieto}, {Souchay}, {Spagna}, {Spoto}, {Stampa}, {Steele},
  {Steidelm{\"u}ller}, {Stephenson}, {Stoev}, {Suess}, {S{\"u}veges}, {Surdej},
  {Szabados}, {Szegedi-Elek}, {Tapiador}, {Taris}, {Tauran}, {Taylor},
  {Teixeira}, {Terrett}, {Tingley}, {Trager}, {Turon}, {Ulla}, {Utrilla},
  {Valentini}, {van Elteren}, {Van Hemelryck}, {van Leeuwen}, {Varadi},
  {Vecchiato}, {Veljanoski}, {Via}, {Vicente}, {Vogt}, {Voss}, {Votruba},
  {Voutsinas}, {Walmsley}, {Weiler}, {Weingrill}, {Werner}, {Wevers},
  {Whitehead}, {Wyrzykowski}, {Yoldas}, {{\v{Z}}erjal}, {Zucker}, {Zurbach},
  {Zwitter}, {Alecu}, {Allen}, {Allende Prieto}, {Amorim},
  {Anglada-Escud{\'e}}, {Arsenijevic}, {Azaz}, {Balm}, {Beck}, {Bernstein},
  {Bigot}, {Bijaoui}, {Blasco}, {Bonfigli}, {Bono}, {Boudreault}, {Bressan},
  {Brown}, {Brunet}, {Bunclark}, {Buonanno}, {Butkevich}, {Carret}, {Carrion},
  {Chemin}, {Ch{\'e}reau}, {Corcione}, {Darmigny}, {de Boer}, {de Teodoro}, {de
  Zeeuw}, {Delle Luche}, {Domingues}, {Dubath}, {Fodor}, {Fr{\'e}zouls},
  {Fries}, {Fustes}, {Fyfe}, {Gallardo}, {Gallegos}, {Gardiol}, {Gebran},
  {Gomboc}, {G{\'o}mez}, {Grux}, {Gueguen}, {Heyrovsky}, {Hoar}, {Iannicola},
  {Isasi Parache}, {Janotto}, {Joliet}, {Jonckheere}, {Keil}, {Kim},
  {Klagyivik}, {Klar}, {Knude}, {Kochukhov}, {Kolka}, {Kos}, {Kutka}, {Lainey},
  {LeBouquin}, {Liu}, {Loreggia}, {Makarov}, {Marseille}, {Martayan},
  {Martinez-Rubi}, {Massart}, {Meynadier}, {Mignot}, {Munari}, {Nguyen},
  {Nordlander}, {Ocvirk}, {O'Flaherty}, {Olias Sanz}, {Ortiz}, {Osorio},
  {Oszkiewicz}, {Ouzounis}, {Palmer}, {Park}, {Pasquato}, {Peltzer}, {Peralta},
  {P{\'e}turaud}, {Pieniluoma}, {Pigozzi}, {Poels}, {Prat}, {Prod'homme},
  {Raison}, {Rebordao}, {Risquez}, {Rocca-Volmerange}, {Rosen}, {Ruiz-Fuertes},
  {Russo}, {Sembay}, {Serraller Vizcaino}, {Short}, {Siebert}, {Silva},
  {Sinachopoulos}, {Slezak}, {Soffel}, {Sosnowska}, {Strai{\v{z}}ys}, {ter
  Linden}, {Terrell}, {Theil}, {Tiede}, {Troisi}, {Tsalmantza}, {Tur},
  {Vaccari}, {Vachier}, {Valles}, {Van Hamme}, {Veltz}, {Virtanen}, {Wallut},
  {Wichmann}, {Wilkinson}, {Ziaeepour}, \& {Zschocke}}]{Prusti:2016}
{Gaia Collaboration}, {Prusti}, T., {de Bruijne}, J.~H.~J., {et~al.} 2016,
  \aap, 595, \dodoi{10.1051/0004-6361/201629272}

\bibitem[{{Gaia Collaboration} {et~al.}(2018{\natexlab{b}}){Gaia
  Collaboration}, {Babusiaux}, {van Leeuwen}, {Barstow}, {Jordi}, {Vallenari},
  {Bossini}, {Bressan}, {Cantat-Gaudin}, {van Leeuwen}, {Brown}, {Prusti}, {de
  Bruijne}, {Bailer-Jones}, {Biermann}, {Evans}, {Eyer}, {Jansen}, {Klioner},
  {Lammers}, {Lindegren}, {Luri}, {Mignard}, {Panem}, {Pourbaix}, {Randich},
  {Sartoretti}, {Siddiqui}, {Soubiran}, {Walton}, {Arenou}, {Bastian},
  {Cropper}, {Drimmel}, {Katz}, {Lattanzi}, {Bakker}, {Cacciari},
  {Casta{\~n}eda}, {Chaoul}, {Cheek}, {De Angeli}, {Fabricius}, {Guerra},
  {Holl}, {Masana}, {Messineo}, {Mowlavi}, {Nienartowicz}, {Panuzzo},
  {Portell}, {Riello}, {Seabroke}, {Tanga}, {Th{\'e}venin}, {Gracia-Abril},
  {Comoretto}, {Garcia-Reinaldos}, {Teyssier}, {Altmann}, {Andrae}, {Audard},
  {Bellas-Velidis}, {Benson}, {Berthier}, {Blomme}, {Burgess}, {Busso},
  {Carry}, {Cellino}, {Clementini}, {Clotet}, {Creevey}, {Davidson}, {De
  Ridder}, {Delchambre}, {Dell'Oro}, {Ducourant},
  {Fern{\'a}ndez-Hern{\'a}ndez}, {Fouesneau}, {Fr{\'e}mat}, {Galluccio},
  {Garc{\'\i}a-Torres}, {Gonz{\'a}lez-N{\'u}{\~n}ez}, {Gonz{\'a}lez-Vidal},
  {Gosset}, {Guy}, {Halbwachs}, {Hambly}, {Harrison}, {Hern{\'a}ndez},
  {Hestroffer}, {Hodgkin}, {Hutton}, {Jasniewicz}, {Jean-Antoine-Piccolo},
  {Jordan}, {Korn}, {Krone-Martins}, {Lanzafame}, {Lebzelter}, {L{\"o}ffler},
  {Manteiga}, {Marrese}, {Mart{\'\i}n-Fleitas}, {Moitinho}, {Mora}, {Muinonen},
  {Osinde}, {Pancino}, {Pauwels}, {Petit}, {Recio-Blanco}, {Richards},
  {Rimoldini}, {Robin}, {Sarro}, {Siopis}, {Smith}, {Sozzetti}, {S{\"u}veges},
  {Torra}, {van Reeven}, {Abbas}, {Abreu Aramburu}, {Accart}, {Aerts},
  {Altavilla}, {{\'A}lvarez}, {Alvarez}, {Alves}, {Anderson}, {Andrei},
  {Anglada Varela}, {Antiche}, {Antoja}, {Arcay}, {Astraatmadja}, {Bach},
  {Baker}, {Balaguer-N{\'u}{\~n}ez}, {Balm}, {Barache}, {Barata}, {Barbato},
  {Barblan}, {Barklem}, {Barrado}, {Barros}, {Bartholom{\'e} Mu{\~n}oz},
  {Bassilana}, {Becciani}, {Bellazzini}, {Berihuete}, {Bertone}, {Bianchi},
  {Bienaym{\'e}}, {Blanco-Cuaresma}, {Boch}, {Boeche}, {Bombrun}, {Borrachero},
  {Bouquillon}, {Bourda}, {Bragaglia}, {Bramante}, {Breddels}, {Brouillet},
  {Br{\"u}semeister}, {Brugaletta}, {Bucciarelli}, {Burlacu}, {Busonero},
  {Butkevich}, {Buzzi}, {Caffau}, {Cancelliere}, {Cannizzaro}, {Carballo},
  {Carlucci}, {Carrasco}, {Casamiquela}, {Castellani}, {Castro-Ginard},
  {Charlot}, {Chemin}, {Chiavassa}, {Cocozza}, {Costigan}, {Cowell}, {Crifo},
  {Crosta}, {Crowley}, {Cuypers}, {Dafonte}, {Damerdji}, {Dapergolas}, {David},
  {David}, {de Laverny}, {De Luise}, {De March}, {de Martino}, {de Souza}, {de
  Torres}, {Debosscher}, {del Pozo}, {Delbo}, {Delgado}, {Delgado}, {Diakite},
  {Diener}, {Distefano}, {Dolding}, {Drazinos}, {Dur{\'a}n}, {Edvardsson},
  {Enke}, {Eriksson}, {Esquej}, {Eynard Bontemps}, {Fabre}, {Fabrizio},
  {Faigler}, {Falc{\~a}o}, {Farr{\`a}s Casas}, {Federici}, {Fedorets},
  {Fernique}, {Figueras}, {Filippi}, {Findeisen}, {Fonti}, {Fraile}, {Fraser},
  {Fr{\'e}zouls}, {Gai}, {Galleti}, {Garabato}, {Garc{\'\i}a-Sedano},
  {Garofalo}, {Garralda}, {Gavel}, {Gavras}, {Gerssen}, {Geyer}, {Giacobbe},
  {Gilmore}, {Girona}, {Giuffrida}, {Glass}, {Gomes}, {Granvik}, {Gueguen},
  {Guerrier}, {Guiraud}, {Guti{\'e}}, {Haigron}, {Hatzidimitriou}, {Hauser},
  {Haywood}, {Heiter}, {Helmi}, {Heu}, {Hilger}, {Hobbs}, {Hofmann}, {Holland},
  {Huckle}, {Hypki}, {Icardi}, {Jan{\ss}en}, {Jevardat de Fombelle}, {Jonker},
  {Juh{\'a}sz}, {Julbe}, {Karampelas}, {Kewley}, {Klar}, {Kochoska}, {Kohley},
  {Kolenberg}, {Kontizas}, {Kontizas}, {Koposov}, {Kordopatis},
  {Kostrzewa-Rutkowska}, {Koubsky}, {Lambert}, {Lanza}, {Lasne}, {Lavigne}, {Le
  Fustec}, {Le Poncin-Lafitte}, {Lebreton}, {Leccia}, {Leclerc},
  {Lecoeur-Taibi}, {Lenhardt}, {Leroux}, {Liao}, {Licata}, {Lindstr{\o}m},
  {Lister}, {Livanou}, {Lobel}, {L{\'o}pez}, {Managau}, {Mann}, {Mantelet},
  {Marchal}, {Marchant}, {Marconi}, {Marinoni}, {Marschalk{\'o}}, {Marshall},
  {Martino}, {Marton}, {Mary}, {Massari}, {Matijevi{\v{c}}}, {Mazeh},
  {McMillan}, {Messina}, {Michalik}, {Millar}, {Molina}, {Molinaro},
  {Moln{\'a}r}, {Montegriffo}, {Mor}, {Morbidelli}, {Morel}, {Morris},
  {Mulone}, {Muraveva}, {Musella}, {Nelemans}, {Nicastro}, {Noval},
  {O'Mullane}, {Ord{\'e}novic}, {Ord{\'o}{\~n}ez-Blanco}, {Osborne}, {Pagani},
  {Pagano}, {Pailler}, {Palacin}, {Palaversa}, {Panahi}, {Pawlak},
  {Piersimoni}, {Pineau}, {Plachy}, {Plum}, {Poggio}, {Poujoulet},
  {Pr{\v{s}}a}, {Pulone}, {Racero}, {Ragaini}, {Rambaux}, {Ramos-Lerate},
  {Regibo}, {Reyl{\'e}}, {Riclet}, {Ripepi}, {Riva}, {Rivard}, {Rixon},
  {Roegiers}, {Roelens}, {Romero-G{\'o}mez}, {Rowell}, {Royer}, {Ruiz-Dern},
  {Sadowski}, {Sagrist{\`a} Sell{\'e}s}, {Sahlmann}, {Salgado}, {Salguero},
  {Sanna}, {Santana-Ros}, {Sarasso}, {Savietto}, {Schultheis}, {Sciacca},
  {Segol}, {Segovia}, {S{\'e}gransan}, {Shih}, {Siltala}, {Silva}, {Smart},
  {Smith}, {Solano}, {Solitro}, {Sordo}, {Soria Nieto}, {Souchay}, {Spagna},
  {Spoto}, {Stampa}, {Steele}, {Steidelm{\"u}ller}, {Stephenson}, {Stoev},
  {Suess}, {Surdej}, {Szabados}, {Szegedi-Elek}, {Tapiador}, {Taris}, {Tauran},
  {Taylor}, {Teixeira}, {Terrett}, {Teyssand ier}, {Thuillot}, {Titarenko},
  {Torra Clotet}, {Turon}, {Ulla}, {Utrilla}, {Uzzi}, {Vaillant}, {Valentini},
  {Valette}, {van Elteren}, {Van Hemelryck}, {Vaschetto}, {Vecchiato},
  {Veljanoski}, {Viala}, {Vicente}, {Vogt}, {von Essen}, {Voss}, {Votruba},
  {Voutsinas}, {Walmsley}, {Weiler}, {Wertz}, {Wevers}, {Wyrzykowski},
  {Yoldas}, {{\v{Z}}erjal}, {Ziaeepour}, {Zorec}, {Zschocke}, {Zucker},
  {Zurbach}, \& {Zwitter}}]{Babusiaux:2018}
{Gaia Collaboration}, {Babusiaux}, C., {van Leeuwen}, F., {et~al.}
  2018{\natexlab{b}}, \aap, 616, A10, \dodoi{10.1051/0004-6361/201832843}

\bibitem[{{Gelman} \& {Rubin}(1992)}]{Gelman:1992}
{Gelman}, A., \& {Rubin}, D.~B. 1992, Statistical Science, 7, 457,
  \dodoi{10.1214/ss/1177011136}

\bibitem[{{Goodman} \& {Weare}(2010)}]{Goodman:2010}
{Goodman}, J., \& {Weare}, J. 2010, Communications in Applied Mathematics and
  Computational Science, 5, 65, \dodoi{10.2140/camcos.2010.5.65}

\bibitem[{{Gossage} {et~al.}(2018){Gossage}, {Conroy}, {Dotter}, {Choi},
  {Rosenfield}, {Cargile}, \& {Dolphin}}]{Gossage:2018}
{Gossage}, S., {Conroy}, C., {Dotter}, A., {et~al.} 2018, \apj, 863, 67,
  \dodoi{10.3847/1538-4357/aad0a0}

\bibitem[{{Green}(2018)}]{dustmaps}
{Green}, G. 2018, The Journal of Open Source Software, 3, 695,
  \dodoi{10.21105/joss.00695}

\bibitem[{{Guhathakurta} \& {Reitzel}(1998)}]{Guhathakurta:1998}
{Guhathakurta}, P., \& {Reitzel}, D.~B. 1998, in Galactic Halos, Vol. 136, 22

\bibitem[{{Hammer} {et~al.}(2015){Hammer}, {Yang}, {Flores}, {Puech}, \&
  {Fouquet}}]{Hammer:2015}
{Hammer}, F., {Yang}, Y.~B., {Flores}, H., {Puech}, M., \& {Fouquet}, S. 2015,
  \apj, 813, 110, \dodoi{10.1088/0004-637X/813/2/110}

\bibitem[{{Harris}(1996)}]{Harris:1996}
{Harris}, W.~E. 1996, \aj, 112, 1487, \dodoi{10.1086/118116}

\bibitem[{{Hernquist}(1990)}]{Hernquist:1990}
{Hernquist}, L. 1990, \apj, 356, 359, \dodoi{10.1086/168845}

\bibitem[{{Hogg} {et~al.}(2010){Hogg}, {Myers}, \& {Bovy}}]{Hogg:2010}
{Hogg}, D.~W., {Myers}, A.~D., \& {Bovy}, J. 2010, \apj, 725, 2166,
  \dodoi{10.1088/0004-637X/725/2/2166}

\bibitem[{{Hunter}(2007)}]{mpl}
{Hunter}, J.~D. 2007, Computing in Science and Engineering, 9, 90,
  \dodoi{10.1109/MCSE.2007.55}

\bibitem[{{Ibata} {et~al.}(1994){Ibata}, {Gilmore}, \& {Irwin}}]{Ibata:1994}
{Ibata}, R.~A., {Gilmore}, G., \& {Irwin}, M.~J. 1994, \nat, 370, 194,
  \dodoi{10.1038/370194a0}

\bibitem[{{Johnston} {et~al.}(2019){Johnston}, {Tkachenko}, {Aerts},
  {Molenberghs}, {Bowman}, {Pedersen}, {Buysschaert}, \&
  {P{\'a}pics}}]{Johnston:2019}
{Johnston}, C., {Tkachenko}, A., {Aerts}, C., {et~al.} 2019, \mnras, 482, 1231,
  \dodoi{10.1093/mnras/sty2671}

\bibitem[{Jones {et~al.}(2001--)Jones, Oliphant, Peterson, {et~al.}}]{scipy}
Jones, E., Oliphant, T., Peterson, P., {et~al.} 2001--, {SciPy}: Open source
  scientific tools for {Python}.
\newblock \url{http://www.scipy.org/}

\bibitem[{{Kalberla} {et~al.}(2010){Kalberla}, {McClure-Griffiths}, {Pisano},
  {Calabretta}, {Ford}, {Lockman}, {Staveley-Smith}, {Kerp}, {Winkel},
  {Murphy}, \& {Newton-McGee}}]{Kalberla:2010}
{Kalberla}, P.~M.~W., {McClure-Griffiths}, N.~M., {Pisano}, D.~J., {et~al.}
  2010, \aap, 521, A17, \dodoi{10.1051/0004-6361/200913979}

\bibitem[{{Kallivayalil} {et~al.}(2006){Kallivayalil}, {van der Marel},
  {Alcock}, {Axelrod}, {Cook}, {Drake}, \& {Geha}}]{Kallivayalil:2006}
{Kallivayalil}, N., {van der Marel}, R.~P., {Alcock}, C., {et~al.} 2006, \apj,
  638, 772, \dodoi{10.1086/498972}

\bibitem[{{Kallivayalil} {et~al.}(2013){Kallivayalil}, {van der Marel},
  {Besla}, {Anderson}, \& {Alcock}}]{Kallivayalil:2013}
{Kallivayalil}, N., {van der Marel}, R.~P., {Besla}, G., {Anderson}, J., \&
  {Alcock}, C. 2013, \apj, 764, 161, \dodoi{10.1088/0004-637X/764/2/161}

\bibitem[{{Kroupa}(2001)}]{Kroupa:2001}
{Kroupa}, P. 2001, \mnras, 322, 231, \dodoi{10.1046/j.1365-8711.2001.04022.x}

\bibitem[{{Laporte} {et~al.}(2019){Laporte}, {Belokurov}, {Koposov}, {Smith},
  \& {Hill}}]{Laporte:2019}
{Laporte}, C. F.~P., {Belokurov}, V., {Koposov}, S.~E., {Smith}, M.~C., \&
  {Hill}, V. 2019, arXiv e-prints, arXiv:1907.10678.
\newblock \doarXiv{1907.10678}

\bibitem[{{Lindegren} {et~al.}(2018){Lindegren}, {Hernandez}, {Bombrun},
  {Klioner}, {Bastian}, {Ramos-Lerate}, {de Torres}, {Steidelmuller},
  {Stephenson}, {Hobbs}, {Lammers}, {Biermann}, {Geyer}, {Hilger}, {Michalik},
  {Stampa}, {McMillan}, {Castaneda}, {Clotet}, {Comoretto}, {Davidson},
  {Fabricius}, {Gracia}, {Hambly}, {Hutton}, {Mora}, {Portell}, {van Leeuwen},
  {Abbas}, {Abreu}, {Altmann}, {Andrei}, {Anglada}, {Balaguer-Nunez},
  {Barache}, {Becciani}, {Bertone}, {Bianchi}, {Bouquillon}, {Bourda},
  {Brusemeister}, {Bucciarelli}, {Busonero}, {Buzzi}, {Cancelliere},
  {Carlucci}, {Charlot}, {Cheek}, {Crosta}, {Crowley}, {de Bruijne}, {de
  Felice}, {Drimmel}, {Esquej}, {Fienga}, {Fraile}, {Gai}, {Garralda},
  {Gonzalez-Vidal}, {Guerra}, {Hauser}, {Hofmann}, {Holl}, {Jordan},
  {Lattanzi}, {Lenhardt}, {Liao}, {Licata}, {Lister}, {Loffler}, {Marchant},
  {Martin-Fleitas}, {Messineo}, {Mignard}, {Morbidelli}, {Poggio}, {Riva},
  {Rowell}, {Salguero}, {Sarasso}, {Sciacca}, {Siddiqui}, {Smart}, {Spagna},
  {Steele}, {Taris}, {Torra}, {van Elteren}, {van Reeven}, \&
  {Vecchiato}}]{Lindegren:2018}
{Lindegren}, L., {Hernandez}, J., {Bombrun}, A., {et~al.} 2018, ArXiv e-prints.
\newblock \doarXiv{1804.09366}

\bibitem[{{Mackey} {et~al.}(2017){Mackey}, {Koposov}, {Da Costa}, {Belokurov},
  {Erkal}, {Fraternali}, {McClure-Griffiths}, \& {Fraser}}]{Mackey:2017}
{Mackey}, A.~D., {Koposov}, S.~E., {Da Costa}, G.~S., {et~al.} 2017, \mnras,
  472, 2975, \dodoi{10.1093/mnras/stx2035}

\bibitem[{{Mackey} {et~al.}(2016){Mackey}, {Koposov}, {Erkal}, {Belokurov}, {Da
  Costa}, \& {G{\'o}mez}}]{Mackey:2016}
{Mackey}, A.~D., {Koposov}, S.~E., {Erkal}, D., {et~al.} 2016, \mnras, 459,
  239, \dodoi{10.1093/mnras/stw497}

\bibitem[{{Ma{\'\i}z Apell{\'a}niz} \& {Weiler}(2018)}]{MAW:2018}
{Ma{\'\i}z Apell{\'a}niz}, J., \& {Weiler}, M. 2018, \aap, 619, A180,
  \dodoi{10.1051/0004-6361/201834051}

\bibitem[{{Majewski} {et~al.}(2003){Majewski}, {Skrutskie}, {Weinberg}, \&
  {Ostheimer}}]{Majewski:2003}
{Majewski}, S.~R., {Skrutskie}, M.~F., {Weinberg}, M.~D., \& {Ostheimer}, J.~C.
  2003, \apj, 599, 1082, \dodoi{10.1086/379504}

\bibitem[{{Mathewson} {et~al.}(1974){Mathewson}, {Cleary}, \&
  {Murray}}]{Mathewson:1974}
{Mathewson}, D.~S., {Cleary}, M.~N., \& {Murray}, J.~D. 1974, \apj, 190, 291,
  \dodoi{10.1086/152875}

\bibitem[{{Mayer} {et~al.}(2006){Mayer}, {Mastropietro}, {Wadsley}, {Stadel},
  \& {Moore}}]{Mayer:2006}
{Mayer}, L., {Mastropietro}, C., {Wadsley}, J., {Stadel}, J., \& {Moore}, B.
  2006, \mnras, 369, 1021, \dodoi{10.1111/j.1365-2966.2006.10403.x}

\bibitem[{{McClure-Griffiths} {et~al.}(2008){McClure-Griffiths},
  {Staveley-Smith}, {Lockman}, {Calabretta}, {Ford}, {Kalberla}, {Murphy},
  {Nakanishi}, \& {Pisano}}]{McClure-Griffiths:2008}
{McClure-Griffiths}, N.~M., {Staveley-Smith}, L., {Lockman}, F.~J., {et~al.}
  2008, \apj, 673, L143, \dodoi{10.1086/528683}

\bibitem[{{McClure-Griffiths} {et~al.}(2009){McClure-Griffiths}, {Pisano},
  {Calabretta}, {Ford}, {Lockman}, {Staveley-Smith}, {Kalberla}, {Bailin},
  {Dedes}, {Janowiecki}, {Gibson}, {Murphy}, {Nakanishi}, \&
  {Newton-McGee}}]{McClure-Griffiths:2009}
{McClure-Griffiths}, N.~M., {Pisano}, D.~J., {Calabretta}, M.~R., {et~al.}
  2009, \apjs, 181, 398, \dodoi{10.1088/0067-0049/181/2/398}

\bibitem[{{McConnachie}(2012)}]{McConnachie:2012}
{McConnachie}, A.~W. 2012, \aj, 144, 4, \dodoi{10.1088/0004-6256/144/1/4}

\bibitem[{{Miyamoto} \& {Nagai}(1975)}]{Miyamoto:1975}
{Miyamoto}, M., \& {Nagai}, R. 1975, \pasj, 27, 533

\bibitem[{{Moni Bidin} {et~al.}(2017){Moni Bidin}, {Casetti-Dinescu}, {Girard},
  {Zhang}, {M{\'e}ndez}, {Vieira}, {Korchagin}, \& {van
  Altena}}]{Moni-Bidin:2017}
{Moni Bidin}, C., {Casetti-Dinescu}, D.~I., {Girard}, T.~M., {et~al.} 2017,
  \mnras, 466, 3077, \dodoi{10.1093/mnras/stw3242}

\bibitem[{{Morton}(2015)}]{Morton:2015}
{Morton}, T.~D. 2015, {isochrones: Stellar model grid package}, Astrophysics
  Source Code Library.
\newblock \doeprint{1503.010}

\bibitem[{{Navarro} {et~al.}(1996){Navarro}, {Frenk}, \&
  {White}}]{Navarro:1996}
{Navarro}, J.~F., {Frenk}, C.~S., \& {White}, S.~D.~M. 1996, \apj, 462, 563,
  \dodoi{10.1086/177173}

\bibitem[{{Nidever} {et~al.}(2008){Nidever}, {Majewski}, \& {Butler
  Burton}}]{Nidever:2008}
{Nidever}, D.~L., {Majewski}, S.~R., \& {Butler Burton}, W. 2008, \apj, 679,
  432, \dodoi{10.1086/587042}

\bibitem[{{Nidever} {et~al.}(2010){Nidever}, {Majewski}, {Butler Burton}, \&
  {Nigra}}]{Nidever:2010}
{Nidever}, D.~L., {Majewski}, S.~R., {Butler Burton}, W., \& {Nigra}, L. 2010,
  \apj, 723, 1618, \dodoi{10.1088/0004-637X/723/2/1618}

\bibitem[{{Nidever} {et~al.}(2017){Nidever}, {Olsen}, {Walker}, {Vivas},
  {Blum}, {Kaleida}, {Choi}, {Conn}, {Gruendl}, {Bell}, {Besla}, {Mu{\~n}oz},
  {Gallart}, {Martin}, {Olszewski}, {Saha}, {Monachesi}, {Monelli}, {de Boer},
  {Johnson}, {Zaritsky}, {Stringfellow}, {van der Marel}, {Cioni}, {Jin},
  {Majewski}, {Martinez-Delgado}, {Monteagudo}, {No{\"e}l}, {Bernard},
  {Kunder}, {Chu}, {Bell}, {Santana}, {Frechem}, {Medina}, {Parkash},
  {Ser{\'o}n Navarrete}, \& {Hayes}}]{Nidever:2017}
{Nidever}, D.~L., {Olsen}, K., {Walker}, A.~R., {et~al.} 2017, \aj, 154, 199,
  \dodoi{10.3847/1538-3881/aa8d1c}

\bibitem[{{No{\"e}l} {et~al.}(2013){No{\"e}l}, {Conn}, {Carrera}, {Read},
  {Rix}, \& {Dolphin}}]{Noel:2013}
{No{\"e}l}, N.~E.~D., {Conn}, B.~C., {Carrera}, R., {et~al.} 2013, \apj, 768,
  109, \dodoi{10.1088/0004-637X/768/2/109}

\bibitem[{{Olsen} {et~al.}(2011){Olsen}, {Zaritsky}, {Blum}, {Boyer}, \&
  {Gordon}}]{Olsen:2011}
{Olsen}, K.~A.~G., {Zaritsky}, D., {Blum}, R.~D., {Boyer}, M.~L., \& {Gordon},
  K.~D. 2011, \apj, 737, 29, \dodoi{10.1088/0004-637X/737/1/29}

\bibitem[{{Pardy} {et~al.}(2018){Pardy}, {D{\textquoteright}Onghia}, \&
  {Fox}}]{Pardy:2018}
{Pardy}, S.~A., {D{\textquoteright}Onghia}, E., \& {Fox}, A.~J. 2018, \apj,
  857, 101, \dodoi{10.3847/1538-4357/aab95b}

\bibitem[{{Paxton} {et~al.}(2011){Paxton}, {Bildsten}, {Dotter}, {Herwig},
  {Lesaffre}, \& {Timmes}}]{Paxton:2011}
{Paxton}, B., {Bildsten}, L., {Dotter}, A., {et~al.} 2011, \apjs, 192, 3,
  \dodoi{10.1088/0067-0049/192/1/3}

\bibitem[{{Paxton} {et~al.}(2013){Paxton}, {Cantiello}, {Arras}, {Bildsten},
  {Brown}, {Dotter}, {Mankovich}, {Montgomery}, {Stello}, {Timmes}, \&
  {Townsend}}]{Paxton:2013}
{Paxton}, B., {Cantiello}, M., {Arras}, P., {et~al.} 2013, \apjs, 208, 4,
  \dodoi{10.1088/0067-0049/208/1/4}

\bibitem[{{Paxton} {et~al.}(2015){Paxton}, {Marchant}, {Schwab}, {Bauer},
  {Bildsten}, {Cantiello}, {Dessart}, {Farmer}, {Hu}, {Langer}, {Townsend},
  {Townsley}, \& {Timmes}}]{Paxton:2015}
{Paxton}, B., {Marchant}, P., {Schwab}, J., {et~al.} 2015, \apjs, 220, 15,
  \dodoi{10.1088/0067-0049/220/1/15}

\bibitem[{P\'erez \& Granger(2007)}]{ipython}
P\'erez, F., \& Granger, B.~E. 2007, Computing in Science and Engineering, 9,
  21, \dodoi{10.1109/MCSE.2007.53}

\bibitem[{{Price-Whelan}(2017)}]{gala}
{Price-Whelan}, A.~M. 2017, The Journal of Open Source Software, 2, 388,
  \dodoi{10.21105/joss.00388}

\bibitem[{{Price-Whelan} \& {Foreman-Mackey}(2017)}]{schwimmbad}
{Price-Whelan}, A.~M., \& {Foreman-Mackey}, D. 2017, The Journal of Open Source
  Software, 2, 357, \dodoi{10.21105/joss.00357}

\bibitem[{{Price-Whelan} {et~al.}(2015){Price-Whelan}, {Johnston}, {Sheffield},
  {Laporte}, \& {Sesar}}]{Price-Whelan:2015}
{Price-Whelan}, A.~M., {Johnston}, K.~V., {Sheffield}, A.~A., {Laporte}, C.
  F.~P., \& {Sesar}, B. 2015, Monthly Notices of the Royal Astronomical
  Society, 452, 676, \dodoi{10.1093/mnras/stv1324}

\bibitem[{{Price-Whelan} {et~al.}(2018){Price-Whelan}, {Hogg}, {Rix}, {De Lee},
  {Majewski}, {Nidever}, {Troup}, {Fern{\'a}ndez-Trincado},
  {Garc{\'\i}a-Hern{\'a}ndez}, {Longa-Pe{\~n}a}, {Nitschelm}, {Sobeck}, \&
  {Zamora}}]{Price-Whelan:2018}
{Price-Whelan}, A.~M., {Hogg}, D.~W., {Rix}, H.-W., {et~al.} 2018, \aj, 156,
  18, \dodoi{10.3847/1538-3881/aac387}

\bibitem[{{Putman} {et~al.}(1998){Putman}, {Gibson}, {Staveley-Smith}, {Banks},
  {Barnes}, {Bhatal}, {Disney}, {Ekers}, {Freeman}, {Haynes}, {Henning},
  {Jerjen}, {Kilborn}, {Koribalski}, {Knezek}, {Malin}, {Mould}, {Oosterloo},
  {Price}, {Ryder}, {Sadler}, {Stewart}, {Stootman}, {Vaile}, {Webster}, \&
  {Wright}}]{Putman:1998}
{Putman}, M.~E., {Gibson}, B.~K., {Staveley-Smith}, L., {et~al.} 1998, \nat,
  394, 752, \dodoi{10.1038/29466}

\bibitem[{{Richter} {et~al.}(2018){Richter}, {Fox}, {Wakker}, {Howk}, {Lehner},
  {Barger}, {D'Onghia}, \& {Lockman}}]{Richter:2018}
{Richter}, P., {Fox}, A.~J., {Wakker}, B.~P., {et~al.} 2018, ArXiv e-prints.
\newblock \doarXiv{1808.09455}

\bibitem[{{Salpeter}(1955)}]{Salpeter:1955}
{Salpeter}, E.~E. 1955, \apj, 121, 161, \dodoi{10.1086/145971}

\bibitem[{{Schlafly} \& {Finkbeiner}(2011)}]{Schlafly:2011}
{Schlafly}, E.~F., \& {Finkbeiner}, D.~P. 2011, \apj, 737, 103,
  \dodoi{10.1088/0004-637X/737/2/103}

\bibitem[{{Schlegel} {et~al.}(1998){Schlegel}, {Finkbeiner}, \&
  {Davis}}]{Schlegel:1998}
{Schlegel}, D.~J., {Finkbeiner}, D.~P., \& {Davis}, M. 1998, \apj, 500, 525,
  \dodoi{10.1086/305772}

\bibitem[{{Smith}(1963)}]{Smith:1963}
{Smith}, G.~P. 1963, Bulletin of the Astronomical Institutes of the
  Netherlands, 17, 203

\bibitem[{{Stark} {et~al.}(2015){Stark}, {Baker}, \& {Kannappan}}]{Stark:2015}
{Stark}, D.~V., {Baker}, A.~D., \& {Kannappan}, S.~J. 2015, \mnras, 446, 1855,
  \dodoi{10.1093/mnras/stu2182}

\bibitem[{{Stetson}(1987)}]{Stetson:1987}
{Stetson}, P.~B. 1987, \pasp, 99, 191, \dodoi{10.1086/131977}

\bibitem[{{Stetson}(1994)}]{Stetson:1994}
---. 1994, \pasp, 106, 250, \dodoi{10.1086/133378}

\bibitem[{{Tepper-Garc{\'{\i}}a} \&
  {Bland-Hawthorn}(2018)}]{Tepper-Garcia:2018}
{Tepper-Garc{\'{\i}}a}, T., \& {Bland-Hawthorn}, J. 2018, \mnras, 478, 5263,
  \dodoi{10.1093/mnras/sty1359}

\bibitem[{{Tepper-Garc{\'\i}a} {et~al.}(2019){Tepper-Garc{\'\i}a},
  {Bland-Hawthorn}, {Pawlowski}, \& {Fritz}}]{Tepper:2019}
{Tepper-Garc{\'\i}a}, T., {Bland-Hawthorn}, J., {Pawlowski}, M.~S., \& {Fritz},
  T.~K. 2019, \mnras, 1575, \dodoi{10.1093/mnras/stz1659}

\bibitem[{{Turner} {et~al.}(2017){Turner}, {Carraro}, \& {Panko}}]{Turner:2017}
{Turner}, D.~G., {Carraro}, G., \& {Panko}, E.~A. 2017, \mnras, 470, 481,
  \dodoi{10.1093/mnras/stx1258}

\bibitem[{{Valdes} {et~al.}(2014){Valdes}, {Gruendl}, \& {DES
  Project}}]{Valdes:2014}
{Valdes}, F., {Gruendl}, R., \& {DES Project}. 2014, in Astronomical Society of
  the Pacific Conference Series, Vol. 485, Astronomical Data Analysis Software
  and Systems XXIII, ed. N.~{Manset} \& P.~{Forshay}, 379

\bibitem[{{Venzmer} {et~al.}(2012){Venzmer}, {Kerp}, \&
  {Kalberla}}]{Venzmer:2012}
{Venzmer}, M.~S., {Kerp}, J., \& {Kalberla}, P.~M.~W. 2012, \aap, 547, A12,
  \dodoi{10.1051/0004-6361/201118677}

\bibitem[{Walt {et~al.}(2011)Walt, Colbert, \& Varoquaux}]{numpy}
Walt, S. v.~d., Colbert, S.~C., \& Varoquaux, G. 2011, Computing in Science and
  Engg., 13, 22, \dodoi{10.1109/MCSE.2011.37}

\bibitem[{{Wolf} {et~al.}(2018){Wolf}, {Onken}, {Luvaul}, {Schmidt}, {Bessell},
  {Chang}, {Da Costa}, {Mackey}, {Martin-Jones}, {Murphy}, {Preston}, {Scalzo},
  {Shao}, {Smillie}, {Tisserand}, {White}, \& {Yuan}}]{Wolf:2018}
{Wolf}, C., {Onken}, C.~A., {Luvaul}, L.~C., {et~al.} 2018, \pasa, 35, e010,
  \dodoi{10.1017/pasa.2018.5}

\bibitem[{{Yang} \& {Tian}(2017)}]{Yang:2017}
{Yang}, W., \& {Tian}, Z. 2017, \apj, 836, 102,
  \dodoi{10.3847/1538-4357/aa5b9d}

\bibitem[{{Zhang} {et~al.}(2017){Zhang}, {Moni Bidin}, {Casetti-Dinescu},
  {M{\'e}ndez}, {Girard}, {Korchagin}, {Vieira}, {van Altena}, \&
  {Zhao}}]{Zhang:2017}
{Zhang}, L., {Moni Bidin}, C., {Casetti-Dinescu}, D.~I., {et~al.} 2017, \apj,
  835, 285, \dodoi{10.3847/1538-4357/835/2/285}

\bibitem[{{Zivick} {et~al.}(2018){Zivick}, {Kallivayalil}, {van der Marel},
  {Besla}, {Linden}, {Koz{\l}owski}, {Fritz}, {Kochanek}, {Anderson}, {Sohn},
  {Geha}, \& {Alcock}}]{Zivick:2018}
{Zivick}, P., {Kallivayalil}, N., {van der Marel}, R.~P., {et~al.} 2018, \apj,
  864, 55, \dodoi{10.3847/1538-4357/aad4b0}

\end{thebibliography}

\end{document}